\documentclass[aps,twocolumn,showpacs,superscriptaddress]{revtex4}
\usepackage{ulem}
\usepackage{graphics,graphicx}
\usepackage{color}
\usepackage{wrapfig}
\usepackage{enumerate}
\usepackage{amssymb,amsmath}
\usepackage{bm}

\begin{document}
\unitlength = 1mm
%~~~~~~~~~~~~~~~~~~~~~~~~~~~~~~~~~~~~~~~~~~~~~~~~~
\title{Quantum Tricriticality in Antiferromagnetic Ising Model with Transverse Field:~\\
A Quantum Monte-Carlo Study}
%~~~~~~~~~~~~~~~~~~~~~~~~~~~~~~~~~~~~~~~~~~~~~~~~~
\author{Yasuyuki Kato} 
\affiliation{Department of Applied Physics, University of Tokyo, Hongo, Bunkyo-ku, Tokyo 113-8656, Japan}
\affiliation{RIKEN Center for Emergent Matter Science (CEMS), Wako, Saitama 351-0198, Japan}
\affiliation{Condensed Matter Theory Laboratory, RIKEN, Wako, Saitama 351-0198, Japan}
\author{Takahiro Misawa}
\affiliation{Department of Applied Physics, University of Tokyo, Hongo, Bunkyo-ku, Tokyo 113-8656, Japan}
%~~~~~~~~~~~~~~~~~~~~~~~~~~~~~~~~~~~~~~~~~~~~~~~~~
\date{\today}
%02.70.Ss Quantum Monte Carlo
%75.30.Kz Magnetic phase transitions
%%%%%%%%%%% %%%%%%%%%%% %%%%%%%%%%% %%%%%%%%%%% %%%%%%%%%%% %%%%%%%%%%%
\begin{abstract}
Quantum tricriticality of a $J_1$-$J_2$ antiferromagnetic Ising model 
on a square lattice is studied using the mean-field (MF) theory, scaling theory,
and the unbiased world-line quantum Monte-Carlo (QMC) method based on the Feynman path integral formula.
The critical exponents of the quantum tricritical point (QTCP) 
and the qualitative phase diagram are obtained from the MF analysis.
By performing the unbiased QMC calculations, 
we provide the numerical evidence for the existence of the QTCP and numerically determine the location of the QTCP in the case of $J_1=J_2$.
From the systematic finite-size scaling analysis,
we conclude that the QTCP is located at $H_{\rm QTCP}/J_1=3.260(2)$ and $\Gamma_{\rm QTCP}/J_1=4.10(5)$.
We also show that the critical exponents of the QTCP are identical to those of the MF theory
because the QTCP in this model is in the upper critical dimension.
The QMC simulations reveal that
unconventional proximity effects of the ferromagnetic susceptibility appear close to the antiferromagnetic QTCP, 
and the proximity effects survive for the conventional quantum critical point.
We suggest that the momentum dependence of the dynamical and static spin structure factors is
useful for identifying the QTCP in experiments. 
\end{abstract}
\maketitle
%%%%%%%%%%% %%%%%%%%%%% %%%%%%%%%%% %%%%%%%%%%% %%%%%%%%%%% %%%%%%%%%%%
% --- --- ---
\section{Introduction}
% --- --- ---

Quantum critical points (QCPs) are often found as a vanishing point of a critical temperature of continuous phase transition by changing external physical parameters such as the magnetic fields and the pressure~\cite{sachdev,stewart,lohneysen,gegenwart}.
It is known that quantum criticalities are governed by the types of symmetry breaking
and the dimensionality as conventional finite-temperature critical points. 
In contrast to the conventional finite-temperature phase transitions,
quantum fluctuations significantly modify the criticality.
Thus, to identify the critical exponents is one of the central issues in the study of the QCPs.
It is also important to reveal the proximity effects of the quantum criticality
because quantum criticality often takes over in a wide parameter space at finite temperature.

According to the quantum-classical mapping~\cite{suzuki,sachdev},
the criticality of the QCP of the symmetry-breaking phase transition in spatial $d$ dimensions
is described by the criticality of $(d+z)$-dimensional classical critical point, 
where $z$ is the dynamical critical exponents.
A typical example of the quantum-classical mapping is the transverse Ising model where the 
quantum phase transition induced by the transverse magnetic field is of $d+1$-dimensional Ising universality class.
The dynamical exponent $z$ can be different from 1 in general.
A prominent example is the so-called magnon BEC transition of magnets near the saturation field where $z=2$~\cite{zapf2014}. 
Another important example is the QCP in itinerant electron systems. 
The theoretical studies using the renormalization-group technique 
have demonstrated that $z=3$ for the ferromagnetic QCP 
while $z=2$ for the antiferromagnetic QCP~\cite{Hertz,Millis}.
This theory indeed successfully explains the non-Fermi 
liquid behavior induced by the QCPs in many materials~\cite{stewart,lohneysen,gegenwart}.
It is also shown that self-consistent renormalization theory
 reproduces the same non-Fermi-liquid behaviors~\cite{SCR,takimoto}.
We note that the phase transitions that are not characterized by 
the conventional symmetry breaking
such as metal-insulator~\cite{ImadaRMP,misawaMQCP} or 
Lifshitz transitions~\cite{Lifshitz1960,YamajiLifshitz} 
do not follow the quantum-classical mapping because they do not have
their classical counterparts.

\begin{figure}[t!]
    \includegraphics[trim = 0 0 0 0,clip,width=8.0cm]{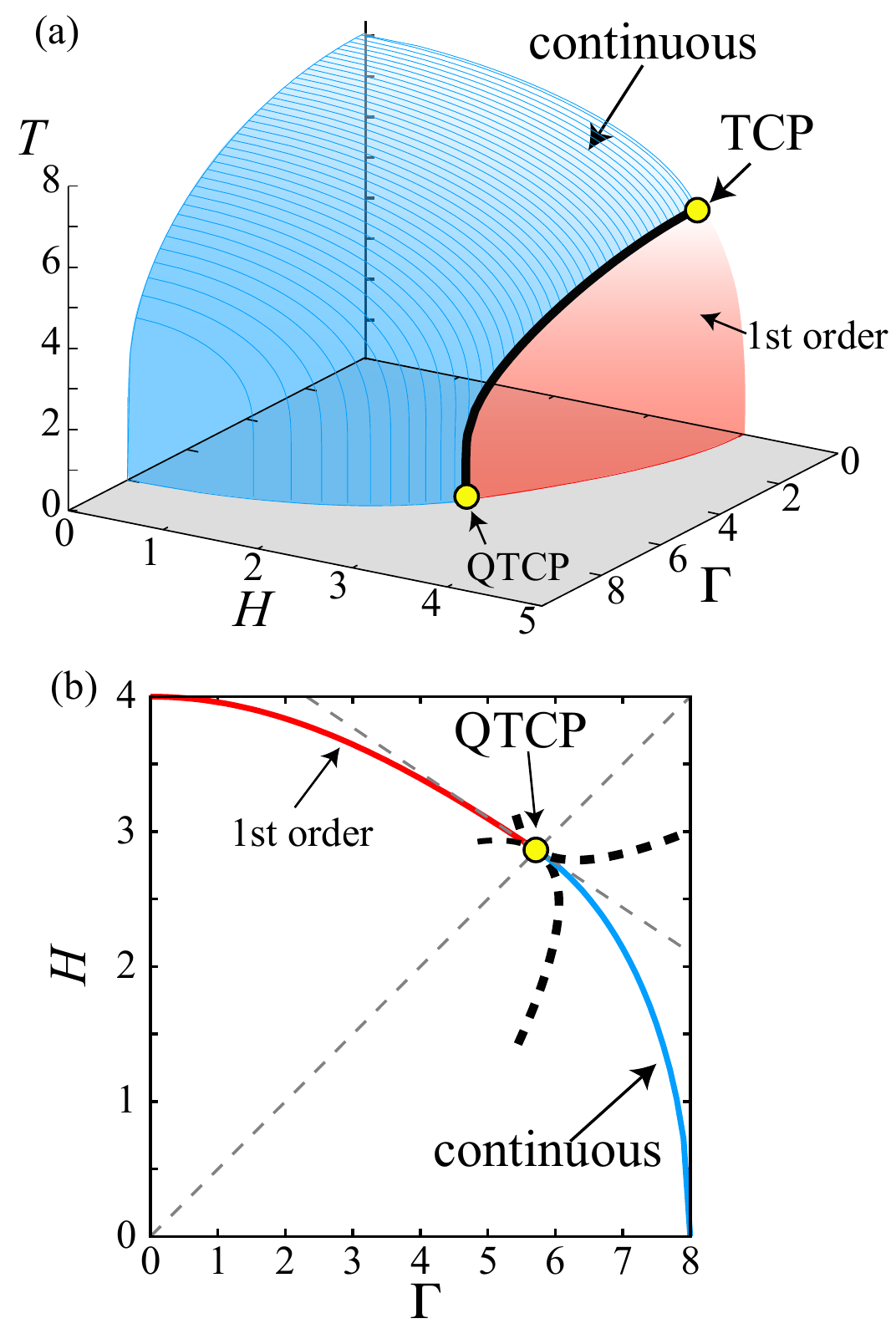}
\caption{(Color online)~(a)~Mean-field phase diagram for $J_1$-$J_2$ model~\eqref{eq:model}.
  We take $J_{1}=J_{2}$ and $J_{1}=1$.
  The left (blue) surface and the right (red) surface represent
  the continuous and first-order phase transition surface, respectively.
  The thick (black) line denotes the tricritical line.
  The continuous phase surface and tricritical line are determined by
  numerically solving the mean-field equations.
  The first-order phase surface is drawn by smoothly interpolating the tricritical line
  and the first-order phase transition line at zero temperature.
  (b) Ground state phase diagram.
  The dashed thin (gray) lines represent
  asymptotic lines for $r=0$ and $u=0$.
  The dashed thick (black) lines represent the crossover curves,
  which are characterized by $u\propto r^{2}$.
  Asymptotic behaviors of $r$ and $u$ around the QTCP
  are given in Eq.~\eqref{eq:ruexp}.
}
\label{Fig:SchematicPhase}
\end{figure}

In contrast to the conventional QCPs,
a proximity effect of first-order quantum phase transitions
induces a quantum tricritical point (QTCP) where a continuous 
phase transition changes into a discontinuous one at zero temperature
[see Fig.~\ref{Fig:SchematicPhase}].
Extending the phase space of the ground state phase diagram towards the field conjugate to the order parameter,
we can see three critical lines (phase boundaries of continuous transition) meet at the QTCP
as we see in finite temperature phase diagrams including a thermal tricritical point (TCP), e.g., phase diagram 
of the spin-1 Blume-Capel model or Blume-Emery-Griffiths model~\cite{TCPreview,cardy1996}.
Therefore, it is expected that more than two different correlation lengths diverge simultaneously; besides, corresponding multi-fluctuations simultaneously diverge at the QTCP.
Several experimental and theoretical works
actually indicate the existence of such QTCPs and
importance of the quantum tricritical fluctuations. 
For example, in heavy-fermion compound YbRh$_{2}$Si$_{2}$~\cite{YRS,gegenwart},
it has been proposed that its unconventional quantum criticalities
are due to a quantum tricriticality~\cite{misawaQTCPletter,misawaQTCPfull}. 
Possibility of the ferromagnetic QTCP has been also discussed in Sr$_{3}$Ru$_{2}$O$_{7}$~\cite{SrRuOQTCP}.
In addition, existence of the antiferromagnetic QTCP has been 
theoretically proposed in iron-based superconductor LaFeAsO~\cite{LaFeAsOQTCP}. 
More recently, it has been shown that the quantum tricritical fluctuations
play key role in stabilizing the superconductivity under magnetic field in URh$_{0.9}$Co$_{0.1}$Ge~\cite{URhCo}.

As is the case of the classical TCP~\cite{TCPreview,nishimori,misawatcp}, 
the criticality of the QTCP is described by the 
$\phi^{6}$ theory instead of the $\phi^{4}$ theory,
and quantum tricriticalities are different from those of
the conventional quantum criticality.
Since the most striking features of the finite-temperature TCP is
the divergence of the concomitant
susceptibility,
it is expected that such a concomitant divergence also occurs
at the QTCP, and its divergence makes 
the proximity effects of the quantum tricriticality 
different from those of the conventional QCP.
For the itinerant electron systems, several phenomenological theories 
have been already proposed for the criticalities 
of the  QTCP~\cite{schmalian,Green,misawaQTCPletter,misawaQTCPfull,Jakub}. 
In the quantum Ising system,
a mean-field calculation of the QTCP~\cite{lukierska1994,BEGQTCP} 
and a renormalization-group study for 
the QTCP in the transverse-like Ising model~\cite{RGQTCP} have been also done.
However, there are few studies that treat
the criticality of the QTCP 
in an unbiased way except for interacting Bosonic systems~\cite{katoQTCP}.

In this paper, 
to clarify the nature of the QTCP,
we perform numerically unbiased large-scale quantum Monte Carlo (QMC) calculations for the transverse field Ising model.
As a result,
we find that the critical temperatures of the TCP are tuned by the transverse and longitudinal
magnetic fields and the QTCP actually appears in the ground state phase diagram.
By performing the systematic finite-size scaling analyses, we clarify the criticality of the QTCP. 
We also examine the momentum dependence of the fluctuations
and the static spin structure factors. In sharp contrast with the ordering
(antiferromagnetic) fluctuations, we find that the concomitant
(ferromagnetic) fluctuations
and the static spin correlations show peculiar momentum dependence. 
This characteristic momentum dependence is a smoking gun for the QTCP.
In addition to that,
we examine the proximity effects of the QTCP
in the paramagnetic phase, which hardly captured by the 
mean-field-type treatment.
As a result,
we find that the ferromagnetic susceptibility has a peak structure, and then
the peak position converges into the QTCP approaching the critical field.
Because the peak structure itself 
still survives for the conventional QCP,
this behavior can be regarded as a remnant of the QTCP.
We will also discuss the relation between the proximity effects and the experimental results. 

This paper is organized as follows:
In Sec.~\ref{sec:model},
we introduce the $J_{1}$-$J_{2}$ model
with the transverse and longitudinal magnetic fields.
We also briefly explain the QMC method.
In Sec.~\ref{sec:MF} we show the results of the 
mean-field calculations and the scaling theory for the QTCP.
In Sec.~\ref{sec:results}, 
we show the results of QMC simulations that are the ground state and finite temperature phase diagrams of the $J_{1}$-$J_{2}$ model, 
the finite-size scaling plots and the finite-temperature and the momentum dependence of the dynamical and static spin structure factors.
We also examine the proximity effects of the QTCP in the paramagnetic region.
Section~\ref{sec:disc} is devoted to summary and discussions.

\section{Model and method}
\label{sec:model}
We consider a simple antiferromagnetic Ising model 
with the external magnetic field on a square lattice
with the periodic boundary condition,
\begin{eqnarray}
  \mathcal{H} &=&
  J_1 \sum_{\langle i,j \rangle} \sigma^z_{i} \sigma^z_{j}
  -J_2 \sum_{\langle\langle i,j \rangle\rangle} \sigma^z_{i} \sigma^z_{j}
  \nonumber\\
  &&-H\sum_{i} \sigma^z_{i} 
  -\Gamma \sum_{i} \sigma^x_{i},
  \label{eq:model}
\end{eqnarray}
where the Pauli matrices $\vec{\sigma}_i$ represent a localized spin at site $i$ ($S=1/2$), 
$J_1$-term ($J_2$-term) represents antiferromagnetic (ferromagnetic) Ising 
interaction between the (next) nearest neighbor spins, 
i.e., $J_1>0$ and $J_2>0$,
and $H$-term ($\Gamma$-term) represents the Zeeman coupling of spins 
to longitudinal (transverse) external magnetic fields.
For simplicity, we use units $\hbar=k_B=a=1$ in this paper where $a$ is the lattice constant.
Especially for the classical case ($\Gamma=0$), similar models have been widely used for analysing metamagneitc phase transitions in highly anisotropic antiferromagnets such as FeCl$_{2}$, and detailed studies have been done in Refs.~\cite{kincaid1975phase,TCPreview}.
We note that 
the MF phase diagram of the model~\eqref{eq:model} has been studied and summarized in Ref.~\cite{lukierska1994}.
However the tricritical line has not been determined precisely.
To make this paper self-contained and to polish the MF phase diagram,
we also perform the MF calculations.

We perform unbiased QMC simulations 
based on the Feynmann path integral formulation~\cite{kawashima2004} to the model~\eqref{eq:model}
using the cluster algorithm invented by Evertz {\it et al.}~\cite{evertz1993}
that is the pioneering method for global-update QMC simulations. 
To avoid the redundancy, we will not explain the cluster algorithm in detail because the application is rather straightforward. 
In the cluster algorithm, the worldlines described 
in $\sigma^z$ basis ($\sigma^z | m_z \rangle = m_z | m_z \rangle$, $m_z \pm 1$) are 
divided into a number of clusters by randomly placing the so-called vertices that 
are inter-site connectors or on-site disconnectors of worldlines. The density of 
vertices are functions of local states and parameters in the Hamiltonian such as $J_1$, $J_2$, and $\Gamma$.
Then, taking into account the longitudinal magnetic field $H$, 
each cluster is independently flipped with probability,
\begin{eqnarray*}
  p=\frac{ e^{-HM_{\rm cluster}}}{ e^{HM_{\rm cluster}}+ e^{-HM_{\rm cluster}} },
\end{eqnarray*}
where $M_{\rm cluster}$ is integral of $m_z$ in the cluster.  
The cluster algorithm is expected to be inefficient where almost all the spins are ferromagnetically aligned with 
the longitudinal magnetic field because $M_{\rm cluster}$ becomes large, and $p$ becomes exponentially small.
This difficulty is eased by antiferromagnetic interaction ($J_1$-term), which makes $M_{\rm cluster}$ smaller, 
and the Zeeman coupling to the transverse magnetic field ($\Gamma$-term), which makes clusters' size smaller.
Indeed, we obtained well converged data as will be shown later.

\section{Mean-Field theory and critical exponents}
\label{sec:MF}
In this section, we derive the critical exponents of the QTCP using the mean-field theory.
It is important to obtain the mean-field exponents because these exponents are exact when the effective dimension $d+z$ is above the upper critical dimension $d_u + z=3$. 
Indeed, we will see the QMC data consistently reproduce the mean-field exponents in Sec.~\ref{sec:results}.
\subsection{Mean-field calculations for $J_{1}$-$J_{2}$ model}
Let us begin with the mean-field analysis of the $J_{1}$-$J_{2}$ model (\ref{eq:model}).
At first, we define the sublattice magnetization as
\begin{equation*}
\langle \sigma_{i}^{z}\rangle=\begin{cases}
                    m_{\rm f}+m_{\rm af},~~~~(i\in A) \\
                    m_{\rm f}-m_{\rm af},~~~~(i\in B) 
                    \end{cases}
\end{equation*}
where $m_{\rm f}$ ($m_{\rm af}$) is the ferromagnetic (antiferromagnetic)
order parameter, and $A$, $B$ represent the sublattice index.
Using the mean-field decoupling,
\begin{equation*}
\sigma_i^z \sigma_j^z
\to \sigma_{i}^{z} \langle \sigma_j^z \rangle
+ \langle \sigma_i^z \rangle \sigma_{j}^{z} 
-\langle \sigma_i^z \rangle \langle \sigma_j^z \rangle,
\end{equation*}
we obtain the mean-field Hamiltonian as
\begin{eqnarray*}
\mathcal{H}_{\rm MF}&=&\frac{N_{\rm s}}{2}
\left[\mathcal{H}^A_{\rm MF}+\mathcal{H}^B_{\rm MF}
- J_{-}m_{\rm f}^2 + J_{+}m_{\rm af}^2
\right],\nonumber\\
\mathcal{H}^{X}_{\rm MF} &=& -H_X\sigma_{i}^{z} -\Gamma \sigma_{i}^{x},
~~~~~ (i\in X)
\end{eqnarray*}
where $N_{\rm s}$ is the number of spins, $X=A$ or $B$,
\begin{eqnarray*}
H_A&=&-J_{-}m_{\rm f} +J_{+}m_{\rm af}+H,\\
H_B&=&-J_{-}m_{\rm f} -J_{+}m_{\rm af}+H,\\
J_{\pm}&=&J_{1}z_{1}\pm J_{2}z_{2},
\end{eqnarray*}
and $z_{1}$ ($z_{2}$) is the coordination number for the (next) nearest-neighbor bonds.
(For concreteness, $z_1=z_2=4$ for the square lattice.)
By diagonalizing $\mathcal{H}^X_{\rm MF}$,
we obtain the eigenvalues for each sublattice as
\begin{equation*}
E^{X}_{\pm}=\pm E^{X}=\pm\sqrt{H_{X}^2+\Gamma^2}.
\end{equation*}
Using $E^X$, the mean-field free energy density is represented as,
\begin{align*}
&f= - \frac{1}{\beta N_{\rm s}} \ln {\rm Tr} \left[ e^{-\beta \mathcal{H}_{\rm MF}} \right] =
-\frac{T}{2}\ln[e^{-\beta E^A }+e^{\beta E^A}]  \\
&-\frac{T}{2}\ln[e^{-\beta E^B}+e^{ \beta E^B}]
-\frac{J_{-}}{2}m_{\rm f}^2+\frac{J_{+}}{2}m_{\rm af}^2,
\end{align*}
where $T$ is the temperature, and $\beta=1/T$ is the inverse temperature.
Minimizing $f$, we obtain the mean-field solutions.
From 
the stationary condition of $f$,
\begin{align*}
  \frac{\partial f}{\partial m_{\rm f}} = 0,~~
  \frac{\partial f}{\partial m_{\rm af}} = 0,
\end{align*}
we obtain the self-consistent equations,
\begin{eqnarray}
m_{\rm af}&=&\frac{1}{2}\Big[\frac{H_A}{E^A}\tanh[\beta E^A]-\frac{H_B}{E^B}\tanh[\beta E^B]\Big], \nonumber \\
m_{\rm f} &=&\frac{1}{2}\Big[\frac{H_A}{E^A}\tanh[\beta E^A]+\frac{H_B}{E^B}\tanh[\beta E^B]\Big] \label{eq:disorder}.
\end{eqnarray}
Solving these self-consistent equations,
we obtain the mean-field phase diagram [see Fig.~\ref{Fig:SchematicPhase}].

Let us consider a simple case $J_{-}=J_{1}z_{1}-J_{2}z_{2}=0$
where the free energy $f$ does not contain $m_{\rm f}$ explicitly, and 
qualitative feature is not different from the case of $J_{-}\neq 0$.
In this case, at $T=0$,
we can easily expand the free energy 
as a function of $m_{\rm af}$ up to sixth order as
\begin{align}
  f=f_{0}+\frac{r}{2}m_{\rm af}^{2}+\frac{u}{4}m_{\rm af}^{4}+\frac{v}{6}m_{\rm af}^6,
  \label{Eq:GL}
\end{align}
where explicit forms of coefficients $f_{0},r,u,v$ are given as,
\begin{align*}
f_{0}&=-\Delta ,\\
r&=\frac{1}{2} J_+ \left(1-\frac{\Gamma^2 J_+}{\Delta^3}\right),\\
u&=\frac{(\Gamma^2-4H^2)\Gamma^2 J^4_+}{8\Delta^7}, \\
v&=\frac{(12\Gamma^2H^2-8H^4-\Gamma^4)\Gamma^2 J_+^6}{16\Delta^{11}},\\
\Delta &= \sqrt{\Gamma^2+H^{2}}.
\end{align*}
The conventional continuous phase transition occurs at $r=0$ when $u>0$,
and the first-order phase transition occurs when $u<0$.
Thus, the location of the QTCP, 
where the continuous phase transition changes into the
first-order phase transition, is determined from $r=u=0$, i.e.,
\begin{align*}
  H_{\rm QTCP}&=\frac{4\sqrt{5}}{25}J_{+},\\
  \Gamma_{\rm QTCP}&=2H_{\rm QTCP}.
\end{align*}
  ($(H_{\rm QTCP}, \Gamma_{\rm QTCP})\simeq(2.86,5.72)$ when $J_1=J_2=1$ as shown in Fig.~\ref{Fig:SchematicPhase}.)
Then, $r$ and $u$ are expanded around the QTCP as
\begin{eqnarray}
  r \simeq \frac{\sqrt{5}}{8}  \delta_\Gamma +\frac{3 \sqrt{5}}{8} \delta_H, ~~
  u \simeq \frac{25 \sqrt{5}}{128} \delta_\Gamma -\frac{ 25 \sqrt{5} }{64} \delta_H, \label{eq:ruexp}
\end{eqnarray}
where $\delta_H \equiv H-H_{\rm QTCP}$, and $\delta_\Gamma \equiv \Gamma-\Gamma_{\rm QTCP}$.

\subsection{Mean-field critical exponents}\label{subsec:mfe}
Here, we discuss the critical exponents of the quantum criticality.
As it is easily understood from the expansion of the free energy in Eq.~(\ref{Eq:GL}), 
the mean-field critical exponents of QTCP are the same as those of finite-temperature TCP. 
We note that the mean-field critical exponents are 
exact above the upper critical dimensions,
which is given by $d_u+z = 3$.
As we will see later, 
the mean-field critical exponents 
are expected to be observed in the two and higher dimensional $J_{1}$-$J_{2}$ model
because $z=1$.
We also show the finite-temperature properties near by the QTCP using the scaling theory.

At the QTCP, both $m_{\rm af}$ and $m_{\rm f}$ are expected to exhibit singularity.
Let us consider the critical exponents regarding $m_{\rm af}$ first.
Supposing the vicinity of the critical point ($m_{\rm af} \ll 1$),
we obtain a simplified self-consistent equation as
\begin{align*} 
  m_{\rm af}[r+um_{\rm af}^{2}+v m_{\rm af}^{4}]=0,
\end{align*} 
from the free-energy expansion in Eq.~(\ref{Eq:GL}).
This equation is easily soluble and  (assuming $m_{\rm af}\neq0, v>0$, and $r<0$),
$m_{\rm af}$ is represented as
\begin{align}
  m_{\rm af}^2=\frac{-u+(u^2-4rv)^{1/2}}{2v}.
\label{Eq:mdag}
\end{align}
For $u=0$, we obtain $m_{\rm af}\sim |r|^{\beta_{t}}, \beta_{t}=1/4$.
This critical exponent is 
nothing but the thermal tricritical exponent.
On the other hand, for $r=0$,
we obtain a different exponent as $m_{\rm af}\sim |u|^{\beta^*_t}, \beta^*_t=1/2$.
In other words, the critical exponents of the QTCP depend on the way of approaching the QTCP in general
because both $r$ and $u$ include $\mathcal{O}(\delta_\Gamma)$ and $\mathcal{O}(\delta_H)$ as shown in Eqs.~\eqref{eq:ruexp}.
These critical behaviours lead to
a scaling relation equation for the order parameter $m_{\rm af}$ as
\begin{align*}
  m_{\rm af}=|r|^{\beta_{t}}\mathcal{M} \left(\frac{|u|}{|r|^{\phi_{t}}} \right),
\end{align*}
where $\mathcal{M}$ is the scaling function, and $\phi_{t}=\beta_t/\beta^*_t=1/2$ is the crossover exponent
(this relation equation can be easily confirmed by Eq.~(\ref{Eq:mdag})). 
Crossover line is defined from the condition $|u|/|r|^{\phi_{t}}\sim \mathcal{O} (1)$, i.e.,
$m_{\rm af}\sim \delta_{(\Gamma H)}^{\beta_t}$ is observed when $|u| \ll |r|^{\phi_{t}}$, while
$m_{\rm af}\sim \delta_{(\Gamma H)}^{\beta^*_t}$ when $|u| \gg |r|^{\phi_{t}}$ where $\delta_{(\Gamma H)} \equiv \sqrt{ \delta_\Gamma^2 + \delta_H^2}$.
Since $\phi_t < 1$, the primary singularity is $m_{\rm af}\sim \delta_{(\Gamma H)}^{\beta_t}$ when approaching the QTCP from generic direction in the phase space.
Only when approaching the QTCP from the special direction with $r=0$ ($\delta_\Gamma = -3 \delta_H$), the primary singularity is $m_{\rm af}\sim \delta_{(\Gamma H)}^{\beta^*_t}$ [see Fig.~\ref{Fig:SchematicPhase}(b)].
Therefore, we will consider only the generic case in this paper.
%

%\begin{align*}
%\frac{H \tanh \left(b \sqrt{G^2+H^2}\right)}{\sqrt{G^2+H^2}} \notag \\
%   -\frac{m^2 \left(H J^2 \text{sech}^2\left(b\sqrt{G^2+H^2}\right) \left(\left(G^2 \left(4 b^2 H^2+3\right)
%   +4 b^2 H^4+3 G^2 \cosh \left(2 b
%   \sqrt{G^2+H^2}\right)\right) \tanh \left(b \sqrt{G^2+H^2}\right)-6 b G^2 \sqrt{G^2+H^2}\right)\right)}{4
%   \left(G^2+H^2\right)^{5/2}}+\frac{H J^4 m^4 \left(\tanh \left(b \sqrt{G^2+H^2}\right) \left(16 b^4 H^8-30 b^2
%   G^6+b \tanh \left(b \sqrt{G^2+H^2}\right) \left(5 G^2 \sqrt{G^2+H^2} \left(4 H^2 \left(4 b^2 G^2-3\right)+16 b^2
%   H^4+9 G^2\right)+2 b \left(G^2+H^2\right) \tanh \left(b \sqrt{G^2+H^2}\right) \left(6 b H^2 \tanh \left(b
%   \sqrt{G^2+H^2}\right) \left(2 b H^2 \left(G^2+H^2\right) \tanh \left(b \sqrt{G^2+H^2}\right)-5 G^2
%   \sqrt{G^2+H^2}\right)-5 \left(G^2 \left(4 b^2 H^4+6 H^2\right)+4 b^2 H^6-3 G^4\right)\right)\right)+G^4 \left(16
%   b^4 H^4+30 b^2 H^2+45\right)+4 G^2 H^2 \left(8 b^4 H^4+15 b^2 H^2-15\right)\right)-5 b G^2 \sqrt{G^2+H^2} \left(4
%   H^2 \left(b^2 G^2-3\right)+4 b^2 H^4+9 G^2\right)\right)}{24 \left(G^2+H^2\right)^{9/2}}+O\left(m^5\right)
%\end{align*}

Next let us consider the critical exponent regarding $m_{\rm f}$.
The singularity of the ferromagnetic order parameter $m_{\rm f}$ around the QTCP is obtained from Eq.~(\ref{eq:disorder}).
Expanding $m_{\rm f}$ with respect to $m_{\rm af}$,
we obtain the relation
\begin{align*}
m_{\rm f} = a_{0}+a_{1} m_{\rm af}^2+a_{2} m_{\rm af}^4+\cdots,
\end{align*}
where $a_{n}$ are constants 
that do not include $m_{\rm f}$ and $m_{\rm af}$.
Associated with the singularity of antiferromagnetic order parameter
 $m_{\rm af}\sim|r|^{\beta_{t}}$,
the singularity of $m_{\rm f}$ is obtained as 
\begin{equation*}
m_{\rm f} -a_0  \sim |r|^{\beta_{2t}}, ~~~ \beta_{2t}=2\beta_{t}=1/2.
\end{equation*}
Besides, the singularity of the ferromagnetic susceptibility is obtained as
\begin{align*}
\chi_{zz}&=\frac{\partial m_{\rm f}}{\partial H} \sim \frac{\partial m_{\rm f}}{\partial r} \sim |r|^{-\gamma_{2t}},\\
\gamma_{2t}&=-2\beta_{t}+1=\frac{1}{2} > 0.
\end{align*}
Therefore, the ferromagnetic susceptibility generally diverges at the QTCP.
Indeed, we will confirm the divergence at the QTCP by performing the numerically unbiased calculations.
By the conventional argument,
we can derive the other critical exponents,
$\delta_{t}$, $\nu_t$, and $\alpha_{t}$, which are defined as
\begin{align*}
m_{\rm af}&\sim |h_s|^{1/\delta_{t}}, \\
\xi&\sim |r|^{-\nu_t}, \\
f_{s}&\sim |r|^{2-\alpha_{t}},
\end{align*}
where $h_s$ is the staggered magnetic field conjugate to the antiferromagnetic order parameter,
$\xi$ is the correlation length associated with the antiferromagnetic order,
and $f_{s}$ is the singular part of the free energy.
In the mean-field theory,
each critical exponent is given by
$\delta_{t}=5$, $\nu_t=1/2$, and $\alpha_{t}=1/2$.
We note that the anomalous dimension $\eta_t$ is zero
within the mean field theory.
Lastly, we summarize the mean-field critical exponents 
for the QTCP in Table.~\ref{Table:Exponents}.
\begin{table}[h]
  \begin{center}
  \begin{tabular}{llllllllll} \hline
                                     & $\alpha_t$    &  $\beta_t$ & $\gamma_t$ & $\delta_t$ & $\beta_{2t}$ & $\gamma_{2t}$  & $\eta_t$ & $\nu_t$          & $\phi_t$  \\    \hline
    Mean field         & 1/2             &  1/4          &      1      & 5   & 1/2          & 1/2            & 0          & 1/2                & 1/2       
    \\ \hline
  \end{tabular}
  \end{center}
\caption{List of mean-field critical exponents for QTCP.
  Mean-filed critical exponents become exact above the upper critical dimensions 
  $d_u+z=3$.}
\label{Table:Exponents}
\end{table}%

\subsection{Scaling theory}
Here, we discuss the finite-temperature properties of the QTCP 
by employing the scaling theory with the finite temperature
 analogue of the finite size scaling hypothesis $\beta/\xi^z \sim \mathcal{O}(1)$ in the vicinity of the QTCP. 
The singular part of the free energy is expressed as
\begin{align*}
f_{s}&\sim|r|^{2-\alpha_{t}}\mathcal{F}
\Big(\frac{u}{|r|^{\phi_{t}}},\frac{|h_s|}{|r|^{\delta_{t}\beta_{t}}},\frac{T}{|r|^{\nu_t z}}\Big), \\
&\sim T^{(2-{\alpha}_{t})/\nu_t z}\tilde{\mathcal{F}}
\Big(\frac{r}{T^{1/\nu_t z}},\frac{u}{T^{\phi_{t}/\nu_t z}},\frac{|h_s|}{T^{\delta_{t}\beta_{t}/\nu_t z}}\Big),
\end{align*}
where $\mathcal{F}$ and $\tilde{\mathcal{F}}$ are scaling functions.
%Using the hyper-scaling relation $2-\alpha_{t}=\nu_t(d+z)$,
We obtain the singularities of $m_{\rm af}$ and $m_{\rm f}$ from $f_s$ as
\begin{align*}
&m_{\rm af}\sim \frac{\partial f_{s}}{\partial h_s} 
\sim |r|^{2-\alpha_{t}-\delta_{t}\beta_{t}} 
\sim T^{(2-\alpha_{t}-\delta_{t}\beta_{t})/\nu_t z}, \\
&m_{\rm f} \sim \frac{\partial f_{s}}{\partial r} 
\sim |r|^{1-\alpha_{t}} 
\sim T^{(1-\alpha_{t})/\nu_t z}.
\end{align*}
The susceptibilities are also
obtained as
\begin{align*}
&\chi_{zz}^{s}=\frac{\partial m_{\rm af}}{\partial h_s} 
\sim |r|^{2-\alpha_{t}-2\delta_{t}\beta_{t}} \sim T^{(2-\alpha_t-2\delta_{t}\beta_{t})/\nu_t z}, \\
&\chi_{zz} \sim \frac{\partial m}{\partial r}
\sim |r|^{-\alpha_{t}} \sim T^{-\alpha_{t}/\nu_t z}.
\end{align*}
Temperature dependence of the specific heat ($C$) is given by
\begin{align*}
C\sim T\frac{\partial^{2} 
f_{s}}{\partial T^{2}} \sim T^{(2-\alpha_{t})/\nu_t z-1} = T^{d/z},
\end{align*}
where we use the hyper-scaling relation ($2-\alpha_{t}=\nu_t(d+z)$) to derive the last relation.
(The specific heat and Sommerfeld constant ($\gamma_{\rm S}=C/T$) are zero at $T=0$,
and does not show $r$ dependence.)
We note that the same temperature dependence is derived by assuming the dispersion of
the low energy excitation proportional to $k^z$, where $k$ represents wavenumber.
Using the mean-field critical exponents and assuming $z=1$,
the temperature dependences of the fluctuations around the two-dimensional QTCP
are summarized as,
\begin{align}
  \chi_{zz}^{s}&\sim T^{-2}, \label{eq:Tchis}\\
  \chi_{zz}&\sim T^{-1},  \label{eq:Tchi}\\
  C&\sim T^{2}.  \label{eq:Tspecific}
\end{align}

We note that the dangerously irrelevant variables generally exist above the upper critical dimensions, and 
the simple scaling argument does not hold and leads incorrect critical exponents~\cite{sachdev,nishimori,kato2010}.
Indeed, this is the reason why the critical exponents of the QTCP in  
itinerant-electron system~\cite{misawaQTCPletter,misawaQTCPfull} are apparently 
inconsistent with the critical exponents derived from the simple scaling argument.
Further detailed calculations on dangerously irrelevant parameters 
 are necessary to derive the correct critical exponents for those cases.
On the other hand, the obtained temperature dependence of the specific heat
$C\propto T^{d/z}$ is expected to be hold even above the upper critical dimensions
because the dangerously irrelevant variables do not affect its criticality~\cite{zulicke} except for the logarithmic corrections.

\section{Results of Quantum Monte Carlo calculations}\label{sec:results}
\subsection{Ground state phase diagram}
\begin{figure}[htb]
  \includegraphics[trim = 0 0 0 0, clip,width=8.cm]{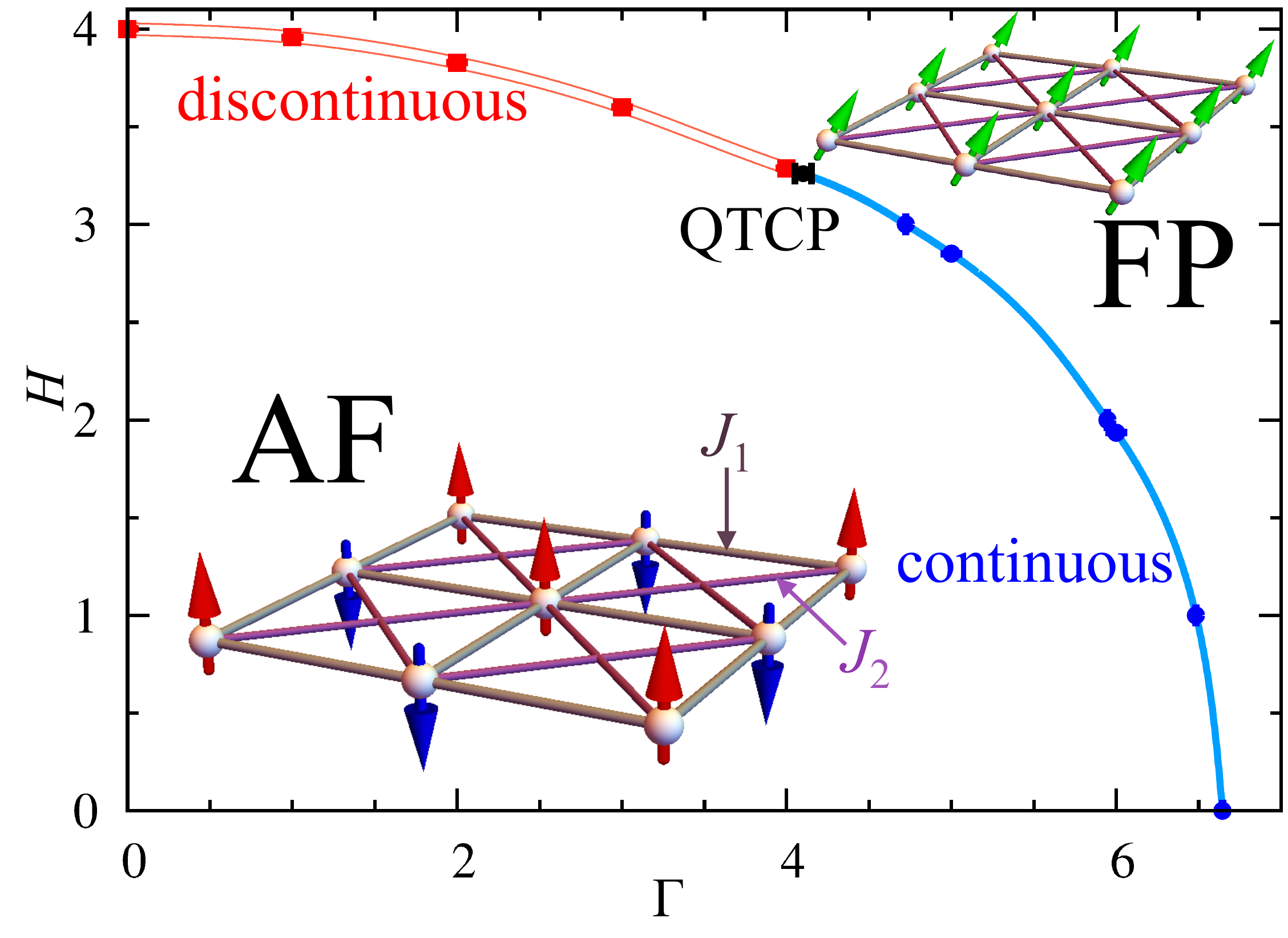} 
  \caption{
    (Color online)
    Ground state phase diagram at $J_1=J_2=1$ obtained by the QMC simulations.
    \label{fig:GSPD}
  }
\end{figure}
Figure~\ref{fig:GSPD} is the ground state phase diagram of the model~\eqref{eq:model} at $J_1=J_2=1$
obtained by the unbiased QMC method.
There is no question that the antiferromagnetic (AF) ordering is stabilized at zero magnetic field at low temperature 
because this system is frustration free.
The model~\eqref{eq:model} without the transverse field ($\Gamma=0$) is a classical Ising spin system.
In this limit, it is well known that the AF phase is stabilized even with finite $H<H^*$.
The ground state is switched into a fully polarized 
(FP) phase from AF phase through the discontinuous phase transition at $H=H^*$.
On the other hand, in another limit where $H=0$, the model is a simple transverse field Ising model,
which exhibits a continuous quantum phase transition to a FP phase.
The universality class is the $2+1$D Ising universality class because the dynamical critical exponent is $z=1$. 
These quantum phase transitions are extended to the region where $H\neq 0$ and $\Gamma \neq 0$ keeping their order of phase transition unchanged,
and then these phase transition lines meet at the QTCP.

\begin{figure}[htb]
  \includegraphics[trim = 0 0 0 0, clip,width=8.cm]{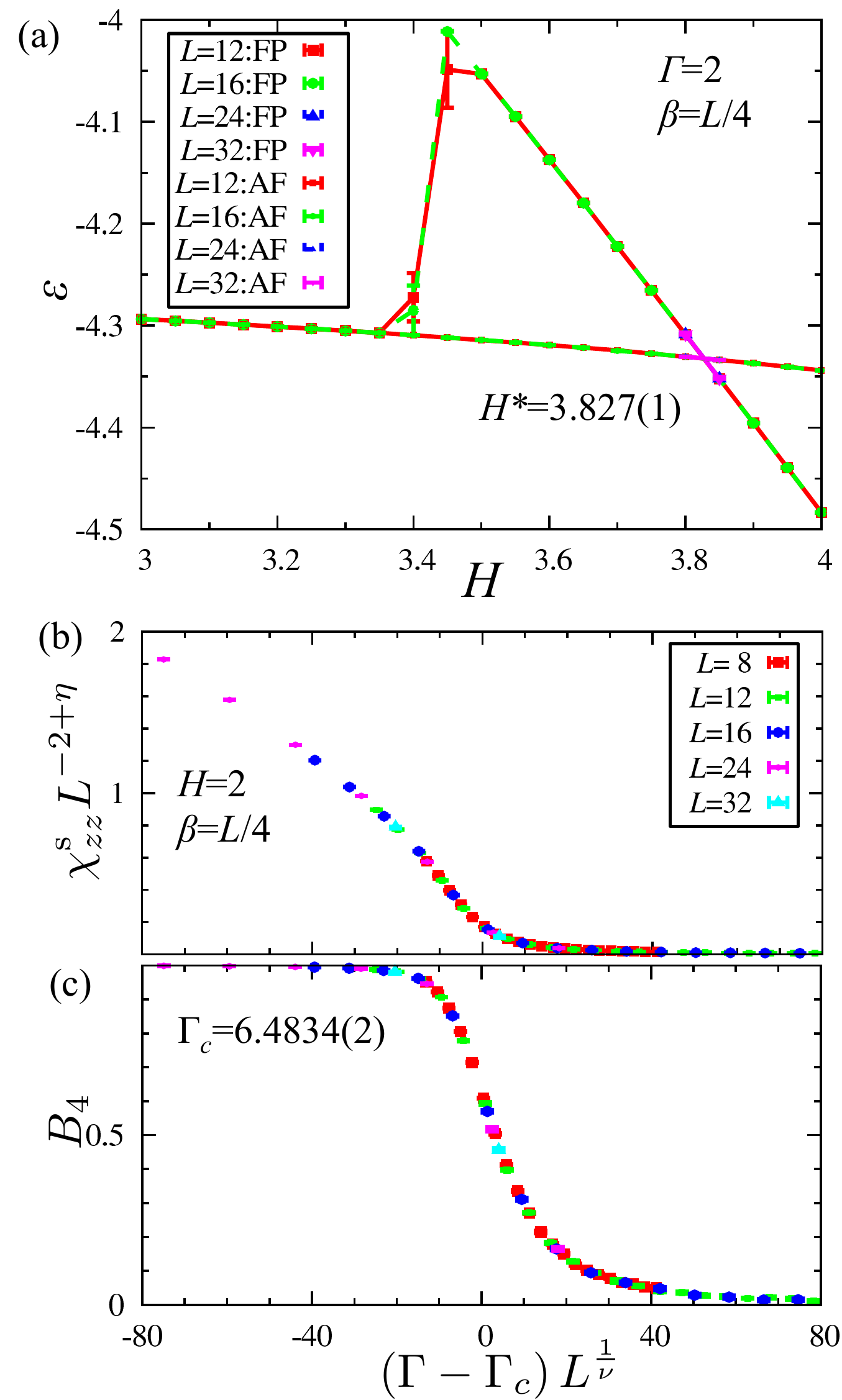} 
  \caption{
    (Color online)
    Determination of the transition points by QMC data.
    We fix $J_1=J_2=1$ and the inverse temperature as $\beta/L=1/4$.    
    (a) Energy level crossing at $\Gamma=2$.
    The QMC simulations are started from either a perfect FP state or a perfect AF state.
    (b,c) Finite size scaling analysis of (b) the staggered magnetic 
   susceptibility $\chi_{zz}^{\rm s}$ and (a) a Binder ratio $B_4$ assuming 
   critical exponents of 3D Ising universality class $\nu=0.6301$, and $\eta=0.0364$ \cite{pelissetto2002}.
    \label{fig:ZTana}
  }
\end{figure}

The  transition points $H^*$ are estimated from the energy-level 
crossing when the transition is discontinuous. In Fig.~\ref{fig:ZTana}(a),
we show an example of energy level crossing at $\Gamma=2$.
We perform  the calculations up to $L=32$ and confirm that the 
finite-size effects are negligibly small.
From this crossing point, we estimate the first-order transition point as
$H^{*}=3.827(1)$ at $\Gamma=2$.

The continuous transition points $H_c$ or $\Gamma_c$ and 
its errors are estimated using the finite size scaling analysis
based on the Bayesian estimate developed by Harada~\cite{harada2011} 
assuming the dynamical critical exponent $z=1$, i.e., 
we increase the system size keeping
the ratio $\beta/L$ constant and
use the 3D Ising critical exponents.
As shown in Figs.~\ref{fig:ZTana}(b,c),
the data of different system sizes are collapsed onto a single curve for both the staggered magnetic susceptibility
\begin{eqnarray*}
 \chi^{\rm s}_{zz} \equiv \frac{\langle M_z  ({\bm Q})^2\rangle}{\beta L^2},
\end{eqnarray*}
and a Binder ratio
\begin{eqnarray*}
B_4 \equiv \frac{1}{2}\left[ 
  3 - \frac{\langle {M_z  ({\bm Q}) }^4 \rangle}{ {\langle {M_z ({\bm Q}) }^2 \rangle}^2 }
\right],
\end{eqnarray*}
where ${\bm Q} = (\pi,\pi)$, 
\begin{eqnarray*}
M_z ({\bm q}) \equiv \int_0^\beta d\tau \sum_{i} \sigma^z_{i}(\tau) e^{- i {\bm q}\cdot{\bm r}_i},
\end{eqnarray*}
and ${\bm r}_i$ is real space coordinate of site $i$.
These well collapsed scaling plots support 
the validity of the assumption $z=1$.

\begin{figure}[htb]
  \includegraphics[trim = 0 0 0 0, clip,width=8.cm]{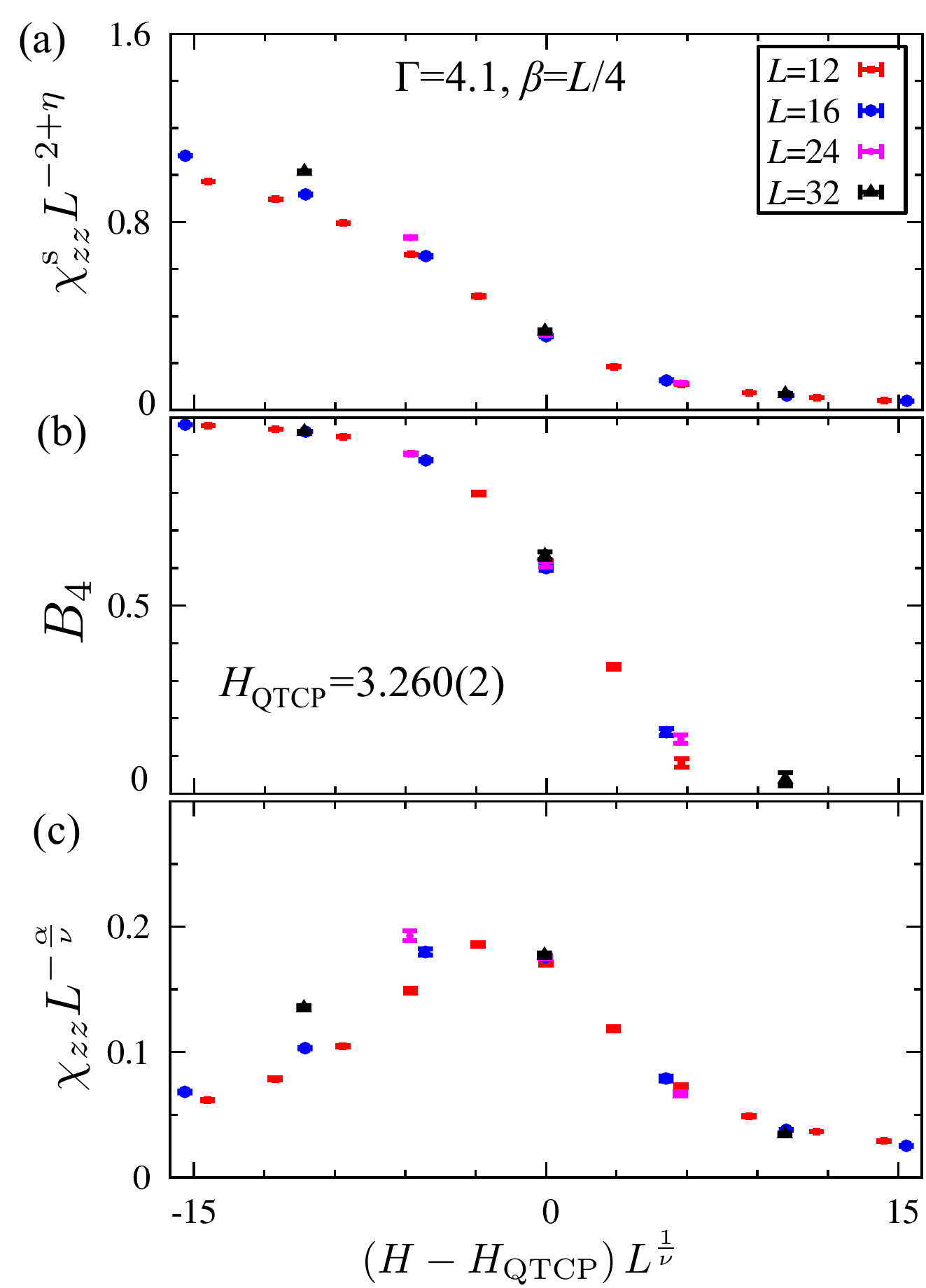} 
  \caption{
    (Color online)
    Finite size scaling analysis at $\Gamma=4.1$ for the QTCP of
    (a) the staggered magnetic susceptibility $\chi_{zz}^{\rm s}$,
    (b) a Binder ratio $B_4$, and
    (c) the uniform magnetic susceptibility $\chi_{zz}$,
    using the critical exponents for the QTCP, $\nu=1/2$, $\eta=0$, and $\alpha=1/2$.
    We fix $J_1=J_2=1$, and the inverse temperature as $\beta/L=1/4$ assuming $z=1$.
    \label{fig:QTCPana}
  }
\end{figure}

From the above analyses, we find that the first-order phase transition
at zero temperature terminates around $\Gamma=4.1$. Thus, as shown in Fig.~\ref{fig:QTCPana}, 
we perform the finite-size scaling analysis for the QTCP.
The critical exponents are expected to be different 
from the 3D Ising critical exponents and are of the mean-field theory
because the upper critical dimension for the QTCP is $d_u=2$ assuming $z=1$.
The position of the QTCP is, indeed, obtained from the finite-size scaling analysis using the exponents derived from the mean-field theory.
The deviation from the single curve in the finite-size scaling plots may be 
  due to rather large step size of $\Gamma$ for searching QTCP (We set the step size as $\Delta \Gamma=0.1$),
  or due to the strong correction to scaling
  i.e., the logarithmic correction due to the dangerous irrelevant variables.
  Another reason of the strong correction to scaling may be the existence of the crossover around the QTCP discussed in Sec.~\ref{subsec:mfe}.
  Actually, a different finite-size scaling form can be derived for $r=0$.
Except for the slight deviations, data are well collapse 
by the quantum tricritical exponents.
This result shows that the QTCP is 
located around $\Gamma_{\rm QTCP} = 4.10(5)$ and $H_{\rm QTCP} =  3.260(2)$ ( $\Gamma_{\rm QTCP}/H_{\rm QTCP}\simeq 1.25$).
For comparison, we show the data for $\Gamma=4.0$ and $\Gamma=4.2$ in Appendix.
We note that the relation $\Gamma_{\rm QTCP}/H_{\rm QTCP}=2$ obtained in the mean-field calculations is strongly
modified by the spatial and quantum fluctuations.
\begin{figure}[htb]
  \includegraphics[trim = 0 0 0 0, clip,width=8.cm]{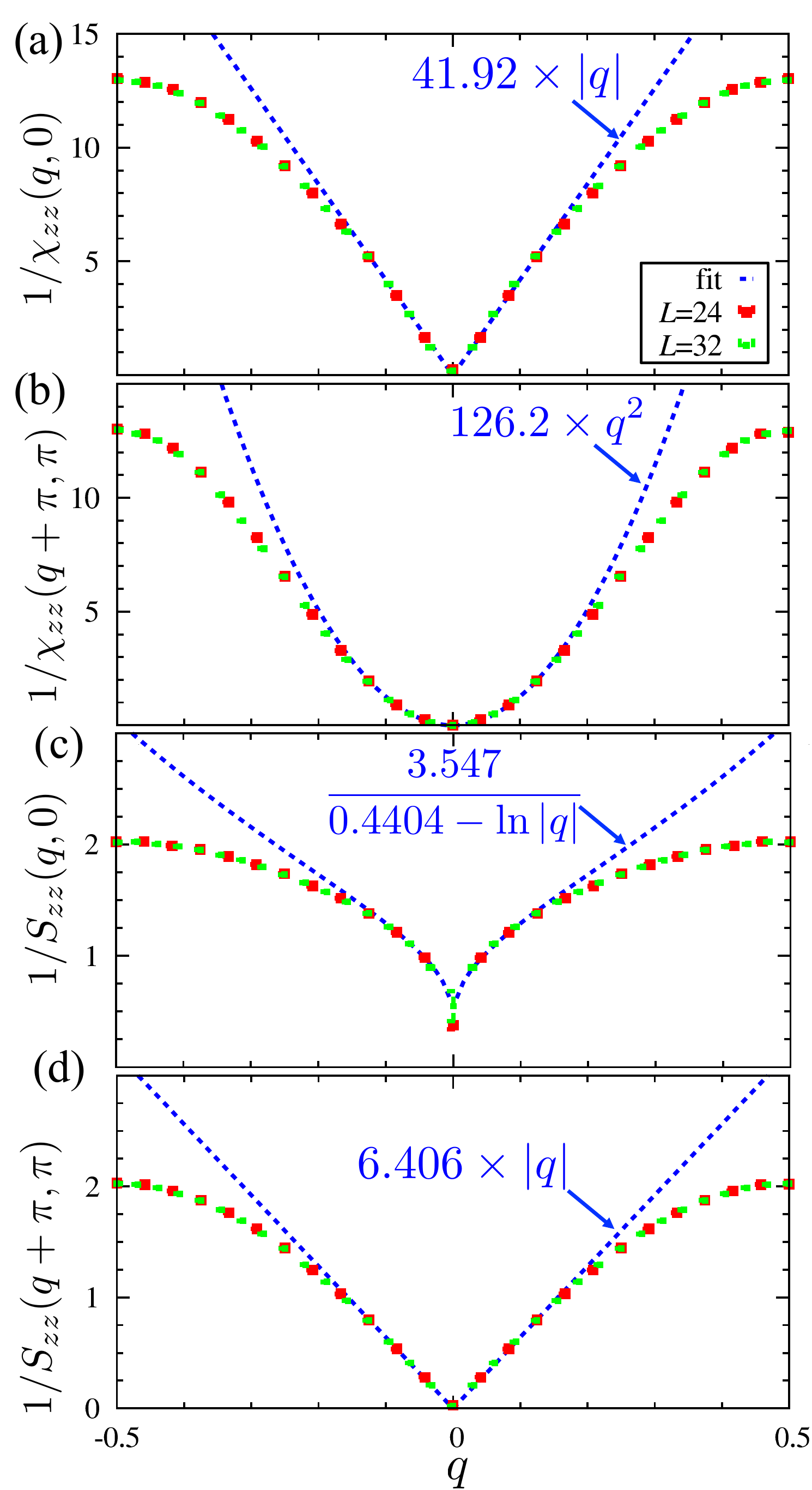} 
  \caption{
    (Color online)
    Low temperature momentum dependence of 
    (a,b) inverse dynamical spin structure factor $\chi^{-1}_{zz}({\bm q})$
    and (c,d) inverse static spin structure factor $S^{-1}_{zz}({\bm q})$
    at $(H_{\rm QTCP},\Gamma_{\rm QTCP})=(4.1,3.26)$ near (a,c) ${\bm q } = 0$
    and (b,d) ${\bm q}={\bm Q}$.
    We fix $J_1=J_2=1$ and the inverse temperature as $\beta/L=1/4$.
    \label{fig:chiq}
  }
\end{figure}

To demonstrate the validity of the estimated value of $H_{\rm QTCP}$ and $\Gamma_{\rm QTCP}$, 
we compute the momentum dependence of the dynamical spin structure factor at zero frequency ($\omega=0$),
\begin{eqnarray*}
\chi_{zz} ( {\bm q} ) &\equiv&
\frac{
\left\langle M_z ({\bm q})^2 \right\rangle
-\left\langle M_z ({\bm q}) \right\rangle^2
}{\beta L^2},\\
&=&\int_0^\beta d\tau \sum_{i} c_{\bm q}(i,\tau),
\end{eqnarray*}
where
\begin{eqnarray*}
c_{\bm q}(i,\tau)&\equiv& \left[
\langle \sigma^z_i (\tau) \sigma^z_0 (0) \rangle
- \langle \sigma^z_i (\tau) \rangle \langle \sigma^z_0 (0) \rangle
\right]  e^{i {\bm q}\cdot({\bm r}_i-{\bm r}_0)},
\end{eqnarray*}
and the static spin structure factor (equal time)
\begin{eqnarray*}
S_{zz} ( {\bm q} ) &\equiv&
\frac{
\left\langle \tilde{M}_z ({\bm q})^2 \right\rangle
-\left\langle \tilde{M}_z ({\bm q}) \right\rangle^2
}{ L^2},\\
&=& \sum_{i} c_{\bm q}(i,0),\\
\tilde{M}_z ({\bm q}) &\equiv& \sum_{i} \sigma^z_{i} e^{- i {\bm q}\cdot{\bm r}_i},
\end{eqnarray*}
at $(H,\Gamma)=(4.1,3.26)$ in low temperature regime [see Fig.~\ref{fig:chiq}].
Note that  $S_{zz}({\bm q})$ 
is observable as the energy integral of the scattering cross section
in the neutron scattering experiments. 
As shown in Fig.~\ref{fig:QTCPana}, 
$\chi_{zz}^{\rm s}=\chi_{zz}({\bm Q})$
and $\chi_{zz}=\chi_{zz}(0)$
are scaled as $\chi_{zz}({\bm Q}) \sim \mathcal{O}(L^2)$, and $\chi_{zz}(0) \sim \mathcal{O}(L)$.
Simple dimensional analysis leads to,
\begin{eqnarray*}
  \chi_{zz} ({\bm q}) &\sim& \frac{1}{|{\bm q}|}, \;\; 
  (|{\bm q}| \ll 1),
  \\
  \chi_{zz} ({\bm q}) &\sim& \frac{1}{({\bm q}-{\bm Q})^2}, \;\; 
  (|{\bm q}-{\bm Q}| \ll 1).
\end{eqnarray*}
We note that the linear $|\bm{q}|$ dependence of 
the concomitant fluctuation has been analytically obtained 
for the exactly solvable model for the thermal TCP~\cite{Emery1975}.
It has been also pointed out that 
simple MF calculations do not reproduce 
the linear $|\bm{q}|$ dependence~\cite{Emery1975,Blume1974}.
Since the imaginary time direction and the real space direction are equally treated ($z=1$),
the correlation function is expected to decay as
\begin{eqnarray*}
  c_{\bm q}(i , \tau )
  \sim
  \left\{
  \begin{array}{ll}
    R^{-2} & ~({\bm q} \simeq 0)\\
    R^{-1} & ~({\bm q} \simeq {\bm Q})
  \end{array}
  \right. ,
\end{eqnarray*}
in the $d+1$-dimensional time space where $R \equiv \sqrt{ (r^x_i -r^x_0)^2+(r^y_i -r^y_0)^2 +\tau^2 }$.
By integrating the correlation function only in the real space with $\tau = 0$, 
the static structure factor is obtained as
$S_{zz}({\bm q}) \sim \mathcal{O}(\log L)$ when $|{\bm q}| \ll 1$,
and $S_{zz}({\bm q}) \sim \mathcal{O}(L)$ when $|{\bm q}-{\bm Q}| \ll 1$.
Again from the simple dimensional analysis,
we obtain the logarithmic and power law singularities of $S_{zz}({\bm q})$ at the QTCP as
\begin{eqnarray*}
  S_{zz} ({\bm q}) &\sim& -\log (|{\bm q}|), \;\; 
  (|{\bm q}| \ll 1),
  \\
  S_{zz} ({\bm q}) &\sim& \frac{1}{|{\bm q}-{\bm Q}|}, \;\; 
  (|{\bm q}-{\bm Q}| \ll 1).
\end{eqnarray*}
Indeed, we confirm that the QMC data show these expected singularities of $\chi_{zz}({\bm q})$ and $S_{zz}({\bm q})$
at the QTCP [see Fig.~\ref{fig:chiq}].
These results strongly suggest the validity of our scaling analysis for the QTCP.

\begin{figure}[htb]
  \includegraphics[trim = 0 0 0 0, clip,width=8.cm]{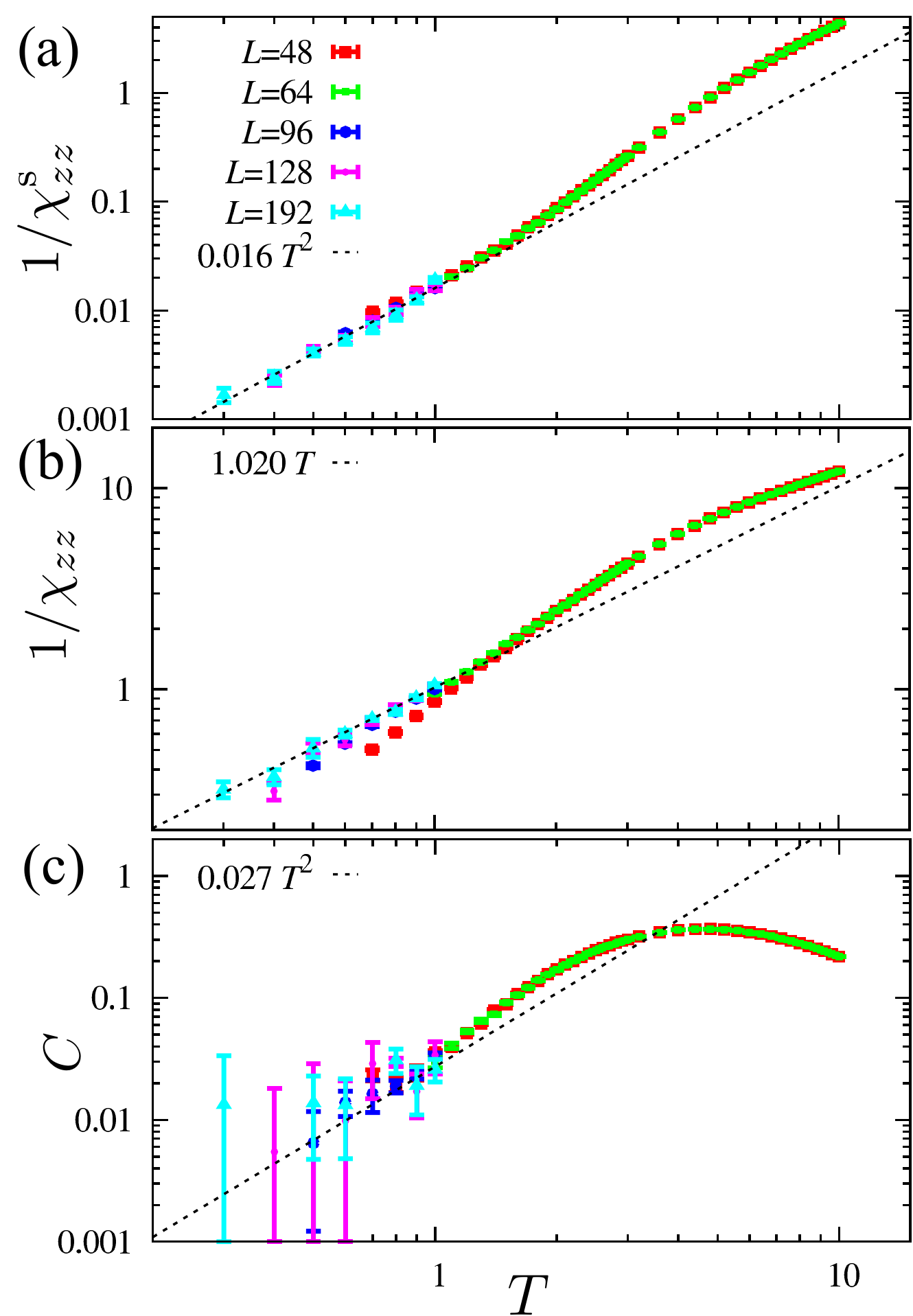} 
  \caption{
    (Color online)
    Temperature dependence of (a) $1/\chi^{\rm s}_{zz}$, (b) $1/\chi_{zz}$ and (c) $C$
    at QTCP, $(\Gamma,H)=(4.1,3.26)$.
    \label{fig:QTCPFT}
  }
\end{figure}

To see the finite-temperature properties of the QTCP,
we compute the temperature dependence of $1/\chi^{\rm s}_{zz}$, $1/\chi_{zz}$ and the specific heat $C$ at 
the QTCP determined by the QMC method 
($\Gamma_{\rm QTCP}=4.1$, $H_{\rm QTCP}=3.26$). 
As shown in Figs.~\ref{fig:QTCPFT},
at sufficient low temperatures and large system sizes,
we confirm that the susceptibilities are well consistent with 
the QTCP exponents derived from the scaling theory, i.e.,
$\chi^{\rm s}_{zz} \sim 1/T^2$, and  $\chi_{zz} \sim 1/T$. 
Although the error bars are relatively large due to the smallness of $C$ at low temperature,
we confirm that the data of specific heat is consistent with $C\sim T^2$,
which is also obtained from the scaling theory. 
All these scaling results indicate that the 2+1 D QTCP exists at $\Gamma_{\rm QTCP}=4.1$ and $H_{\rm QTCP}=3.26$.

\subsection{Finite temperature phase diagrams}
\begin{figure}[htb]
  \includegraphics[trim = 0 0 0 0, clip,width=8.cm]{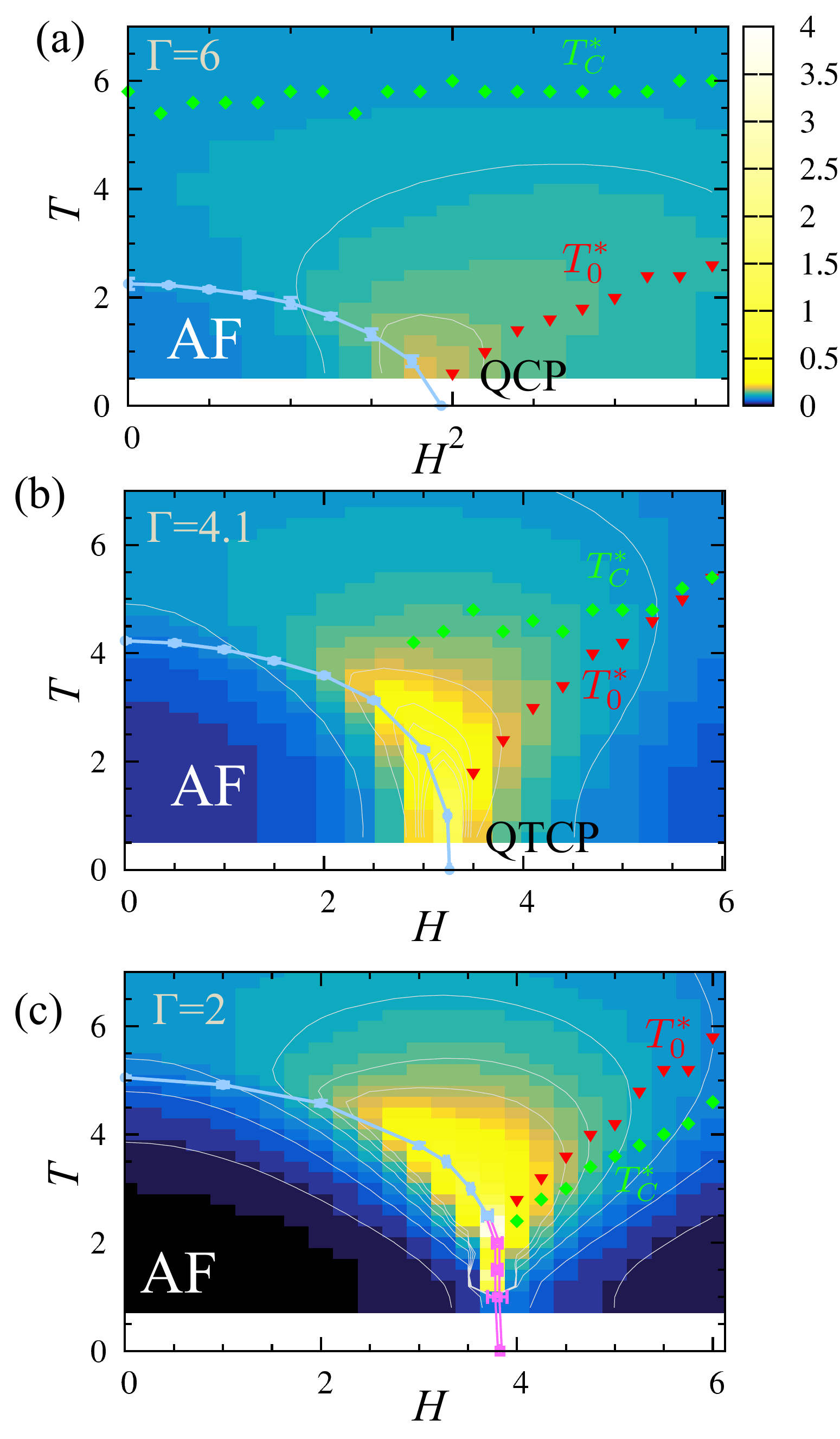} 
  \caption{
    (Color online)
    Finite-temperature phase diagrams at (a) $\Gamma=6$, (b) $\Gamma=4.1$ and (c) $\Gamma=2$,
      where (a) the conventional QCP, (b) the QTCP, and (c) the discontinuous quantum phase transition point
      exist at $T=0$, respectively.
      The solid thick lines (light blue) and double line (pink) show the continuous 
      and the discontinuous phase transition points, respectively.
      $T^*_0$ and $T^*_C$ show the positions of the broad peaks of $\chi_{zz}$ and $C$, respectively.
      The background intensity plots represent $\chi_{zz}$ computed with $L=16$,
      and the thinner lines represent their contours.     
    \label{fig:FTPDs}
  }
\end{figure}

Figures~\ref{fig:FTPDs} show the finite-temperature 
phase diagrams at $\Gamma=$6, 4.1, and 2 where
the quantum phase transition is a generic continuous 
one in the 3D Ising universality class, 
a continuous one with the tricriticality, 
and a discontinuous one, respectively.
The positions of discontinuous transition are determined from 
the discontinuous jumps of the magnetization ($m_x$ and $m_z$). 
On the other hand, the positions of continuous 
transition are determined from the finite-size scaling analysis
of the staggered magnetic susceptibility $\chi^{\rm s}_{zz}$ and the Binder ratio $B_4$ with 
critical exponents of the 2D Ising universality class ($\nu=1$ and $\eta=1/4$). 

\begin{figure}[htb]
  \includegraphics[trim = 0 0 0 0, clip,width=8.cm]{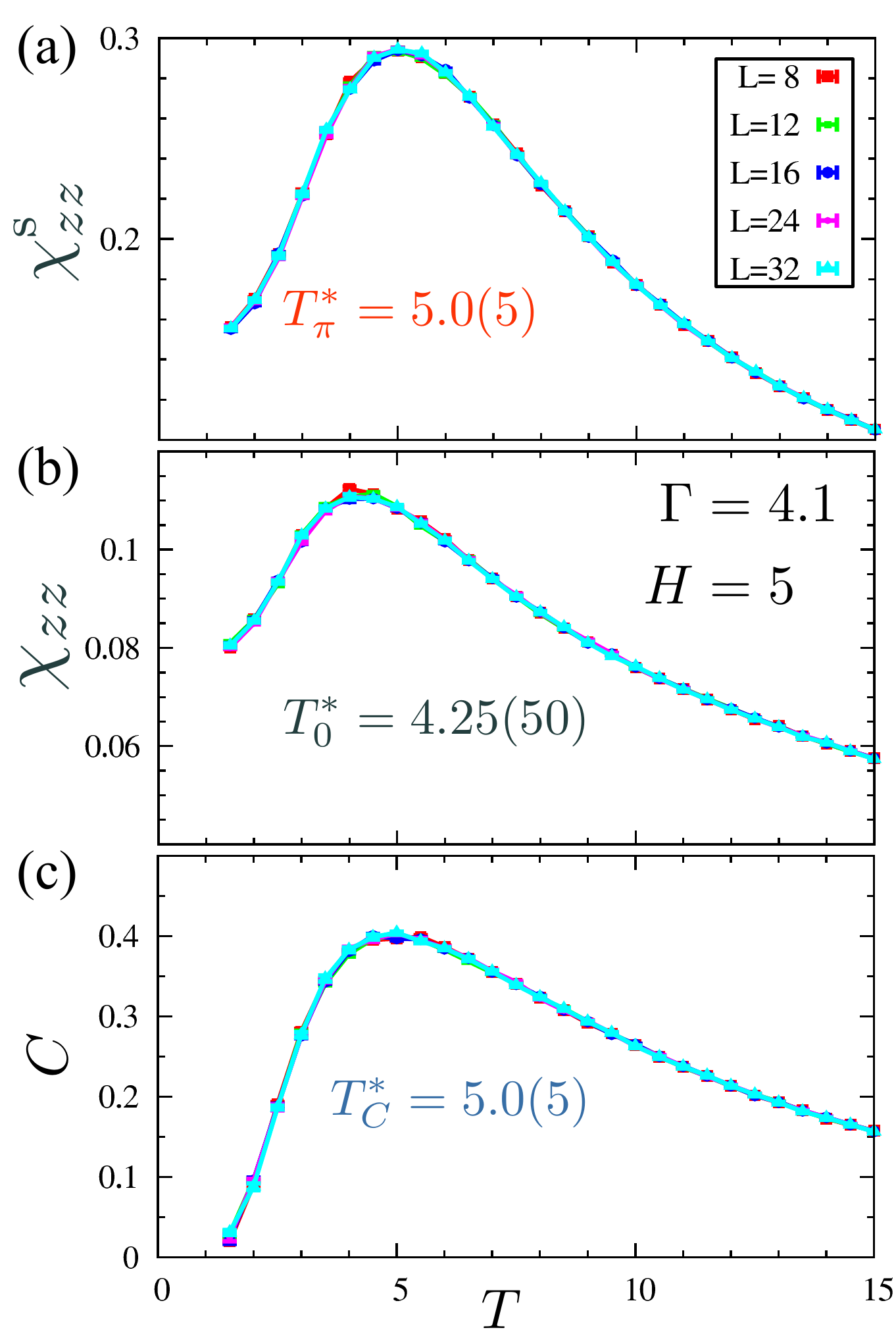} 
  \caption{
    (Color online)
    Temperature dependence of (a) $\chi^{\rm s}_{zz}$, (b) $\chi_{zz}$ and (c) $C$
    at $(\Gamma,H)=(4.1,5)$.
    \label{fig:QTCPPROX}
  }
\end{figure}

In the phase diagrams, we display the 
positions of broad peaks of 
$\chi_{zz}$ and $C$ in paramagnetic phase.
It is well known that the magnetic susceptibilities exhibit 
a broad peak as a proximity effect near a finite-temperature tricritical point 
(e.g., \cite{de1995}).
Indeed, we confirm such a proximity effect in the case of finite-temperature tricritical point in Fig.~\ref{fig:FTPDs}(c):
The both $T^*$s converge on the tricritical point, and the closer $H$ is to the tricritical point, the sharper the peaks are.   
The proximity effects for QTCP exists as well as those of the thermal tricritical point.
We show an example of the proximity effect around the QTCP in Fig.~\ref{fig:QTCPPROX}.
Only difference is that the broad peak of specific heat does not converge to the QTCP and stay at higher temperature.
The reason is simply because the specific heat is zero at $T=0$, and does not diverge at the QTCP.
In other words, the weaker the first-order quantum phase transition is, the weaker the proximity effect of specific heat is.
In the case of the conventional QCP (Fig.~\ref{fig:FTPDs}(a)),
$\chi_{zz}$ exhibits similar broad peak structure,
and $T^*_0$ seems to converge into the QCP.
However, $\chi_{zz}$ does not show divergence at the QCP and 
remains rather small value unlike the QTCP.

%Proximity effect

\section{Discussions and Conclusions}\label{sec:disc}

In conclusion,
we study the $J_1-J_2$ antiferromagnetic 
Ising model with both the longitudinal and the transverse magnetic fields by 
the MF theory, the scaling theory, and the unbiased large-scale QMC calculations.
In the MF theory,
we show that the critical temperature of the TCP can be tuned by the longitudinal and the transverse magnetic fields,
and the QTCP appears at $(\Gamma_{\rm QTCP},H_{\rm QTCP})=(8\sqrt{5}/25J_{+}, 4\sqrt{5}/25J_{+})$ when $J_- = 0$. 
We also clarify the singularity of physical quantities 
associated with the QTCP using the Ginzburg-Landau expansion.
We summarize the critical exponents for the QTCP and 
complete the phase diagram 
in the case of $J_1=J_2$ by the MF analysis. 
Especially we show that the uniform magnetic susceptibility 
$\chi_{zz}$ that is not the ordering but the concomitant 
susceptibility diverges at the QTCP unlike the generic case of QCP.
Using the scaling theory, we also clarify the temperature dependence 
of physical quantities around the QTCP.

By performing the QMC calculations, we obtain
the numerically unbiased phase diagram in the case of $J_1=J_2$.
The QTCP is found at $H_{\rm QTCP}=3.260(2)$ and 
$\Gamma_{\rm QTCP}=4.10(5)$ in our finite-size scaling analysis.
We also examine the momentum dependence 
of the dynamical and static spin structure factors.
All the obtained results are consistent with the expected QTCP singularities.
This consistency strongly supports validity of the scaling analysis.

Furthermore, we examine the temperature dependence of the
antiferromagnetic and ferromagnetic fluctuations around the QTCP and
confirm that the concomitant divergence of the 
ferromagnetic fluctuation occurs at the antiferromagnetic QTCP.
We show that
this divergence induces the characteristic crossover in the 
paramagnetic region around the QTCP; the ferromagnetic susceptibility
has a peak at $T_{0}^{*}$ [see Fig.~\ref{fig:FTPDs}].
We note that the peak structures, which are remnants
of the QTCP, survive for the conventional QCP as shown in Fig~\ref{fig:FTPDs}, 
although the ferromagnetic susceptibility does not diverge at the QCP.
We note that appearance of peak structures of the 
ferromagnetic susceptibility are observed around the antiferromagnetic
QCP in YbRh$_{2}$(Si$_{0.95}$Ge$_{0.05}$)$_{2}$~\cite{YRS} 
and the peak structures may be the remnant of the QTCP.

Lastly, we discuss the experimental identification of the QTCP. 
Recently, anomalous divergent behaviors of the ferromagnetic fluctuations
have been found in several materials.
For example, in YbRh$_{2}$Si$_{2}$, the diverging behaviors of the ferromagnetic fluctuations~\cite{YRS,gegenwart}
have been observed around the antiferromagnetic QCP. 
Furthermore, unconventional divergent behaviors of the ferromagnetic fluctuations have been also observed in YbAlB$_{4}$~\cite{YbAlB42008,YbAlB42011} 
and in a quasi crystal Au$_{51}$Al$_{34}$Yb$_{15}$~\cite{deguchi} 
although any clear symmetry breaking phase transition or QCP have not been found in these materials. 
Several theories such as the valence quantum criticalities~\cite{valence} and the critical nodal metal~\cite{nodal}
have been proposed for explaining the unconventional divergent behaviors of the ferromagnetic fluctuations.
In these theories, although the mechanism of the diverging behaviors of the ferromagnetic fluctuation are different,
it is common that the diverging fluctuations are the critical fluctuations, i.e., the ordering fluctuations.
In contrast to them,
the quantum tricriticality induces the divergence of the concomitant
fluctuation whose momentum dependence is different 
from that of the ordering fluctuation as shown in Fig.~\ref{fig:chiq}.
Therefore, by examining whether the momentum dependence of 
the dynamical and static spin-structure factors near ${\bm q}=0$ show $\chi_{zz}({\bm q}) \sim 1/|{\bm q}|$ 
and $S_{zz}({\bm q}) \sim -1/\log |{\bm q}|$ or not,
it is possible to conclude whether the quantum tricriticality governs 
those unconventional quantum criticalities or not.
Further experimental investigation along this direction will 
reveal the nature of the unconventional quantum criticalities.
It is also an intriguing issue
how the divergence of the concomitant
susceptibility affects the nature of 
the superconductivity observed in YbAlB$_{4}$~\cite{YbAlB42008} and URh$_{1-x}$Co$_{x}$Ge~\cite{URhCo}.

% --- --- ---
\begin{acknowledgements} 
The authors thank Y.~Motome for fruitful discussions and comments. 
They also thank the organizers of the workshop 
on theoretical studies of strongly correlated electron systems in Wakayama in 2014, 
where the early stage of this work started.
This work was supported by JSPS KAKENHI Grant No.\,26800199. 
Numerical calculations were conducted using RICC and HOKUSAI-GW.
\end{acknowledgements} 
% --- --- ---

\appendix*
\section{Finite-size scaling analysis for QTCP}
\begin{figure*}[htb]
  \includegraphics[trim = 0 0 0 0, clip,width=14.cm]{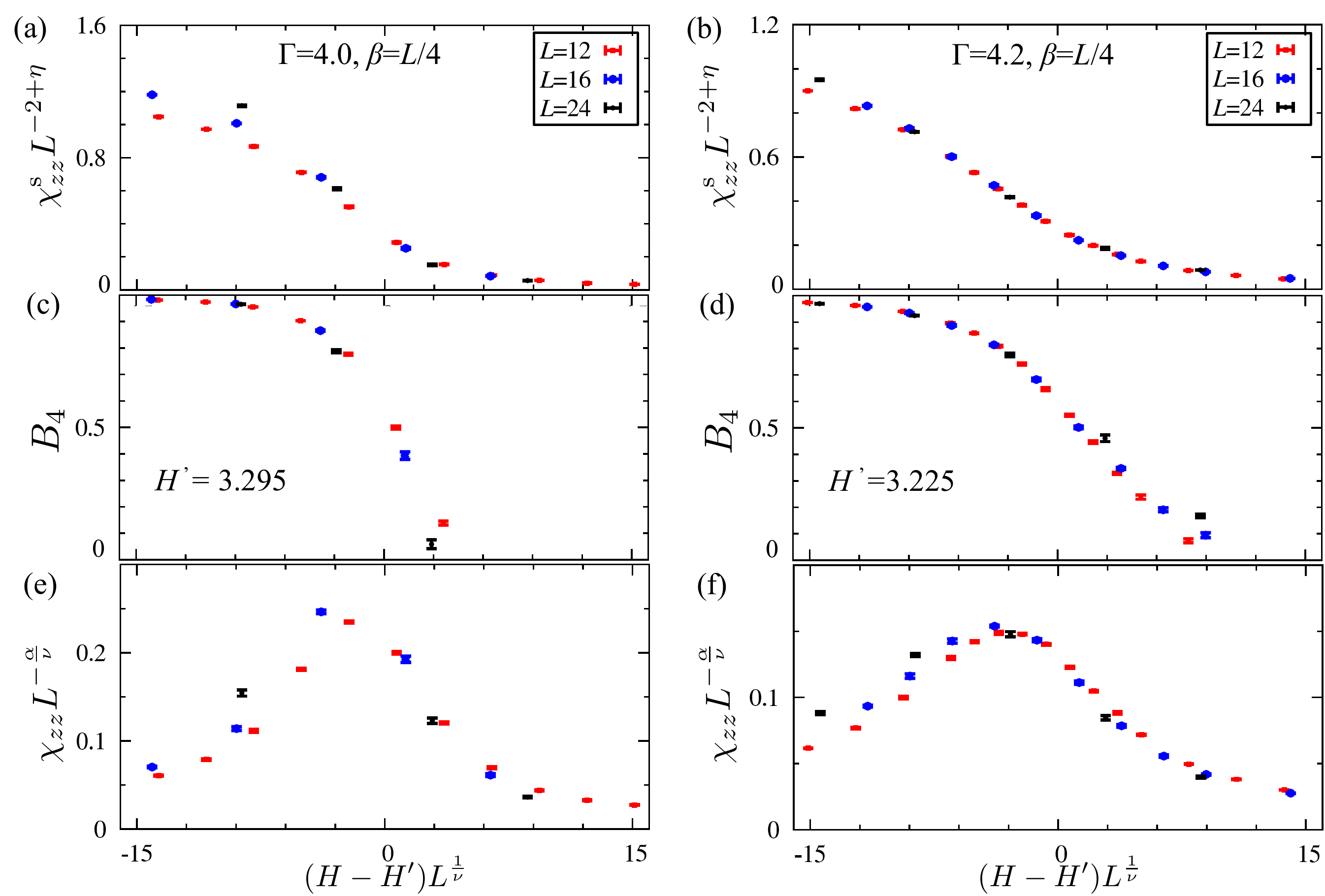} 
  \caption{
    (Color online)
    Finite size scaling analysis at (a,c,e) $\Gamma=4.0$ and (b,d,f) $\Gamma=4.2$ of
    (a,b) the staggered magnetic susceptibility $\chi_{zz}^{\rm s}$,
    (c,d) a Binder ratio $B_4$, and
    (e,f) the uniform magnetic susceptibility $\chi_{zz}$,
    using the critical exponents for the QTCP, $\nu=1/2$, $\eta=0$, and $\alpha=1/2$.
    We fix $J_1=J_2=1$, and the inverse temperature as $\beta/L=1/4$ assuming $z=1$.
    The tuning variable $H'=3.295$ and $H'=3.225$ 
    give the best scaling plot at $\Gamma=4.0$ and $\Gamma=4.2$, respectively.
    \label{fig:QTCPana_app}
  }
\end{figure*}

In the main text, we show the results of the finite-size scaling
at $\Gamma=4.1$, and conclude that $\Gamma_{\rm QTCP}=4.10(5)$.
Figures~\ref{fig:QTCPana_app}(a-f) show the results of 
the finite-size scaling analysis at $\Gamma=4.0$ and $\Gamma=4.2$.
The finite-size scaling plot of $\chi_{zz}$ is sensitive to the deviation from the QTCP
while those of $\chi^{\rm s}_{zz}$ and $B_4$ are insensitive.
In both cases, the data do not show the monotonic convergence with increasing $L$.

%\break
\clearpage

\bibliographystyle{apsrev}
\bibliography{manuscript}

\begin{thebibliography}{46}
\expandafter\ifx\csname natexlab\endcsname\relax\def\natexlab#1{#1}\fi
\expandafter\ifx\csname bibnamefont\endcsname\relax
  \def\bibnamefont#1{#1}\fi
\expandafter\ifx\csname bibfnamefont\endcsname\relax
  \def\bibfnamefont#1{#1}\fi
\expandafter\ifx\csname citenamefont\endcsname\relax
  \def\citenamefont#1{#1}\fi
\expandafter\ifx\csname url\endcsname\relax
  \def\url#1{\texttt{#1}}\fi
\expandafter\ifx\csname urlprefix\endcsname\relax\def\urlprefix{URL }\fi
\providecommand{\bibinfo}[2]{#2}
\providecommand{\eprint}[2][]{\url{#2}}

\bibitem[{\citenamefont{Sachdev}(2007)}]{sachdev}
\bibinfo{author}{\bibfnamefont{S.}~\bibnamefont{Sachdev}},
  \emph{\bibinfo{title}{Quantum phase transitions}} (\bibinfo{publisher}{Wiley
  Online Library}, \bibinfo{year}{2007}).

\bibitem[{\citenamefont{Stewart}(2001)}]{stewart}
\bibinfo{author}{\bibfnamefont{G.~R.} \bibnamefont{Stewart}},
  \bibinfo{journal}{Rev. Mod. Phys.} \textbf{\bibinfo{volume}{73}},
  \bibinfo{pages}{797} (\bibinfo{year}{2001}).

\bibitem[{\citenamefont{L\"ohneysen et~al.}(2007)\citenamefont{L\"ohneysen,
  Rosch, Vojta, and W\"olfle}}]{lohneysen}
\bibinfo{author}{\bibfnamefont{H.~v.} \bibnamefont{L\"ohneysen}},
  \bibinfo{author}{\bibfnamefont{A.}~\bibnamefont{Rosch}},
  \bibinfo{author}{\bibfnamefont{M.}~\bibnamefont{Vojta}}, \bibnamefont{and}
  \bibinfo{author}{\bibfnamefont{P.}~\bibnamefont{W\"olfle}},
  \bibinfo{journal}{Rev. Mod. Phys.} \textbf{\bibinfo{volume}{79}},
  \bibinfo{pages}{1015} (\bibinfo{year}{2007}).

\bibitem[{\citenamefont{Gegenwart et~al.}(2008)\citenamefont{Gegenwart, Si, and
  Steglich}}]{gegenwart}
\bibinfo{author}{\bibfnamefont{P.}~\bibnamefont{Gegenwart}},
  \bibinfo{author}{\bibfnamefont{Q.}~\bibnamefont{Si}}, \bibnamefont{and}
  \bibinfo{author}{\bibfnamefont{F.}~\bibnamefont{Steglich}},
  \bibinfo{journal}{Nat. Phys.} \textbf{\bibinfo{volume}{4}},
  \bibinfo{pages}{186} (\bibinfo{year}{2008}).

\bibitem[{\citenamefont{Suzuki}(1976)}]{suzuki}
\bibinfo{author}{\bibfnamefont{M.}~\bibnamefont{Suzuki}},
  \bibinfo{journal}{Progr. Theor. Phys.} \textbf{\bibinfo{volume}{56}},
  \bibinfo{pages}{1454} (\bibinfo{year}{1976}).

\bibitem[{\citenamefont{Zapf et~al.}(2014)\citenamefont{Zapf, Jaime, and
  Batista}}]{zapf2014}
\bibinfo{author}{\bibfnamefont{V.}~\bibnamefont{Zapf}},
  \bibinfo{author}{\bibfnamefont{M.}~\bibnamefont{Jaime}}, \bibnamefont{and}
  \bibinfo{author}{\bibfnamefont{C.~D.} \bibnamefont{Batista}},
  \bibinfo{journal}{Rev. Mod. Phys.} \textbf{\bibinfo{volume}{86}},
  \bibinfo{pages}{563} (\bibinfo{year}{2014}).

\bibitem[{\citenamefont{Hertz}(1976)}]{Hertz}
\bibinfo{author}{\bibfnamefont{J.~A.} \bibnamefont{Hertz}},
  \bibinfo{journal}{Phys. Rev. B} \textbf{\bibinfo{volume}{14}},
  \bibinfo{pages}{1165} (\bibinfo{year}{1976}).

\bibitem[{\citenamefont{Millis}(1993)}]{Millis}
\bibinfo{author}{\bibfnamefont{A.~J.} \bibnamefont{Millis}},
  \bibinfo{journal}{Phys. Rev. B} \textbf{\bibinfo{volume}{48}},
  \bibinfo{pages}{7183} (\bibinfo{year}{1993}).

\bibitem[{\citenamefont{Moriya}(1985)}]{SCR}
\bibinfo{author}{\bibfnamefont{T.}~\bibnamefont{Moriya}},
  \emph{\bibinfo{title}{Spin fluctuations in itinerant electron \\magnetism}},
  vol.~\bibinfo{volume}{56} (\bibinfo{publisher}{Springer-Verlag Berlin},
  \bibinfo{year}{1985}).

\bibitem[{\citenamefont{Moriya and Takimoto}(1995)}]{takimoto}
\bibinfo{author}{\bibfnamefont{T.}~\bibnamefont{Moriya}} \bibnamefont{and}
  \bibinfo{author}{\bibfnamefont{T.}~\bibnamefont{Takimoto}},
  \bibinfo{journal}{J. Phys. Soc. Jpn.} \textbf{\bibinfo{volume}{64}},
  \bibinfo{pages}{960} (\bibinfo{year}{1995}).

\bibitem[{\citenamefont{Imada et~al.}(1998)\citenamefont{Imada, Fujimori, and
  Tokura}}]{ImadaRMP}
\bibinfo{author}{\bibfnamefont{M.}~\bibnamefont{Imada}},
  \bibinfo{author}{\bibfnamefont{A.}~\bibnamefont{Fujimori}}, \bibnamefont{and}
  \bibinfo{author}{\bibfnamefont{Y.}~\bibnamefont{Tokura}},
  \bibinfo{journal}{Rev. Mod. Phys.} \textbf{\bibinfo{volume}{70}},
  \bibinfo{pages}{1039} (\bibinfo{year}{1998}).

\bibitem[{\citenamefont{Misawa and Imada}(2007)}]{misawaMQCP}
\bibinfo{author}{\bibfnamefont{T.}~\bibnamefont{Misawa}} \bibnamefont{and}
  \bibinfo{author}{\bibfnamefont{M.}~\bibnamefont{Imada}},
  \bibinfo{journal}{Phys. Rev. B} \textbf{\bibinfo{volume}{75}},
  \bibinfo{pages}{115121} (\bibinfo{year}{2007}).

\bibitem[{\citenamefont{Lifshitz}(1960)}]{Lifshitz1960}
\bibinfo{author}{\bibfnamefont{I.~M.} \bibnamefont{Lifshitz}},
  \bibinfo{journal}{Sov. Phys. JETP} \textbf{\bibinfo{volume}{11}},
  \bibinfo{pages}{1130} (\bibinfo{year}{1960}).

\bibitem[{\citenamefont{Yamaji et~al.}(2006)\citenamefont{Yamaji, Misawa, and
  Imada}}]{YamajiLifshitz}
\bibinfo{author}{\bibfnamefont{Y.}~\bibnamefont{Yamaji}},
  \bibinfo{author}{\bibfnamefont{T.}~\bibnamefont{Misawa}}, \bibnamefont{and}
  \bibinfo{author}{\bibfnamefont{M.}~\bibnamefont{Imada}}, \bibinfo{journal}{J.
  Phys. Soc. Jpn.} \textbf{\bibinfo{volume}{75}}, \bibinfo{pages}{094719}
  (\bibinfo{year}{2006}).

\bibitem[{\citenamefont{Lawrie and Sarbach}(1984)}]{TCPreview}
\bibinfo{author}{\bibfnamefont{I.~D.} \bibnamefont{Lawrie}} \bibnamefont{and}
  \bibinfo{author}{\bibfnamefont{S.}~\bibnamefont{Sarbach}},
  \emph{\bibinfo{title}{Phase Transition and Critical \\ Phenomena edited by C.
  Domb and J. L. Lebowitz}}, vol.~\bibinfo{volume}{9}
  (\bibinfo{publisher}{Academic Press, London}, \bibinfo{year}{1984}).

\bibitem[{\citenamefont{Cardy}(1996)}]{cardy1996}
\bibinfo{author}{\bibfnamefont{J.}~\bibnamefont{Cardy}},
  \emph{\bibinfo{title}{Scaling and renormalization in statistical \\
  physics}}, vol.~\bibinfo{volume}{5} (\bibinfo{publisher}{Cambridge university
  press}, \bibinfo{year}{1996}).

\bibitem[{\citenamefont{Gegenwart et~al.}(2005)\citenamefont{Gegenwart,
  Custers, Tokiwa, Geibel, and Steglich}}]{YRS}
\bibinfo{author}{\bibfnamefont{P.}~\bibnamefont{Gegenwart}},
  \bibinfo{author}{\bibfnamefont{J.}~\bibnamefont{Custers}},
  \bibinfo{author}{\bibfnamefont{Y.}~\bibnamefont{Tokiwa}},
  \bibinfo{author}{\bibfnamefont{C.}~\bibnamefont{Geibel}}, \bibnamefont{and}
  \bibinfo{author}{\bibfnamefont{F.}~\bibnamefont{Steglich}},
  \bibinfo{journal}{Phys. Rev. Lett.} \textbf{\bibinfo{volume}{94}},
  \bibinfo{pages}{076402} (\bibinfo{year}{2005}).

\bibitem[{\citenamefont{Misawa et~al.}(2008)\citenamefont{Misawa, Yamaji, and
  Imada}}]{misawaQTCPletter}
\bibinfo{author}{\bibfnamefont{T.}~\bibnamefont{Misawa}},
  \bibinfo{author}{\bibfnamefont{Y.}~\bibnamefont{Yamaji}}, \bibnamefont{and}
  \bibinfo{author}{\bibfnamefont{M.}~\bibnamefont{Imada}}, \bibinfo{journal}{J.
  Phys. Soc. Jpn.} \textbf{\bibinfo{volume}{77}}, \bibinfo{pages}{093712}
  (\bibinfo{year}{2008}).

\bibitem[{\citenamefont{Misawa et~al.}(2009)\citenamefont{Misawa, Yamaji, and
  Imada}}]{misawaQTCPfull}
\bibinfo{author}{\bibfnamefont{T.}~\bibnamefont{Misawa}},
  \bibinfo{author}{\bibfnamefont{Y.}~\bibnamefont{Yamaji}}, \bibnamefont{and}
  \bibinfo{author}{\bibfnamefont{M.}~\bibnamefont{Imada}}, \bibinfo{journal}{J.
  Phys. Soc. Jpn.} \textbf{\bibinfo{volume}{78}}, \bibinfo{pages}{084707}
  (\bibinfo{year}{2009}).

\bibitem[{\citenamefont{Green et~al.}(2005{\natexlab{a}})\citenamefont{Green,
  Grigera, Borzi, Mackenzie, Perry, and Simons}}]{SrRuOQTCP}
\bibinfo{author}{\bibfnamefont{A.~G.} \bibnamefont{Green}},
  \bibinfo{author}{\bibfnamefont{S.~A.} \bibnamefont{Grigera}},
  \bibinfo{author}{\bibfnamefont{R.~A.} \bibnamefont{Borzi}},
  \bibinfo{author}{\bibfnamefont{A.~P.} \bibnamefont{Mackenzie}},
  \bibinfo{author}{\bibfnamefont{R.~S.} \bibnamefont{Perry}}, \bibnamefont{and}
  \bibinfo{author}{\bibfnamefont{B.~D.} \bibnamefont{Simons}},
  \bibinfo{journal}{Phys. Rev. Lett.} \textbf{\bibinfo{volume}{95}},
  \bibinfo{pages}{086402} (\bibinfo{year}{2005}{\natexlab{a}}).

\bibitem[{\citenamefont{Giovannetti et~al.}(2011)\citenamefont{Giovannetti,
  Ortix, Marsman, Capone, van~den Brink, and Lorenzana}}]{LaFeAsOQTCP}
\bibinfo{author}{\bibfnamefont{G.}~\bibnamefont{Giovannetti}},
  \bibinfo{author}{\bibfnamefont{C.}~\bibnamefont{Ortix}},
  \bibinfo{author}{\bibfnamefont{M.}~\bibnamefont{Marsman}},
  \bibinfo{author}{\bibfnamefont{M.}~\bibnamefont{Capone}},
  \bibinfo{author}{\bibfnamefont{J.}~\bibnamefont{van~den Brink}},
  \bibnamefont{and}
  \bibinfo{author}{\bibfnamefont{J.}~\bibnamefont{Lorenzana}},
  \bibinfo{journal}{Nat. Commun.} \textbf{\bibinfo{volume}{2}},
  \bibinfo{pages}{398} (\bibinfo{year}{2011}).

\bibitem[{\citenamefont{Tokunaga et~al.}(2015)\citenamefont{Tokunaga, Aoki,
  Mayaffre, Kr\"amer, Julien, Berthier, Horvati\ifmmode~\acute{c}\else
  \'{c}\fi{}, Sakai, Kambe, and Araki}}]{URhCo}
\bibinfo{author}{\bibfnamefont{Y.}~\bibnamefont{Tokunaga}},
  \bibinfo{author}{\bibfnamefont{D.}~\bibnamefont{Aoki}},
  \bibinfo{author}{\bibfnamefont{H.}~\bibnamefont{Mayaffre}},
  \bibinfo{author}{\bibfnamefont{S.}~\bibnamefont{Kr\"amer}},
  \bibinfo{author}{\bibfnamefont{M.-H.} \bibnamefont{Julien}},
  \bibinfo{author}{\bibfnamefont{C.}~\bibnamefont{Berthier}},
  \bibinfo{author}{\bibfnamefont{M.}~\bibnamefont{Horvati\ifmmode~\acute{c}\else
  \'{c}\fi{}}}, \bibinfo{author}{\bibfnamefont{H.}~\bibnamefont{Sakai}},
  \bibinfo{author}{\bibfnamefont{S.}~\bibnamefont{Kambe}}, \bibnamefont{and}
  \bibinfo{author}{\bibfnamefont{S.}~\bibnamefont{Araki}},
  \bibinfo{journal}{Phys. Rev. Lett.} \textbf{\bibinfo{volume}{114}},
  \bibinfo{pages}{216401} (\bibinfo{year}{2015}).

\bibitem[{\citenamefont{Nishimori and Ortiz}(2010)}]{nishimori}
\bibinfo{author}{\bibfnamefont{H.}~\bibnamefont{Nishimori}} \bibnamefont{and}
  \bibinfo{author}{\bibfnamefont{G.}~\bibnamefont{Ortiz}},
  \emph{\bibinfo{title}{Elements of Phase \\ Transitions and Critical
  Phenomena}} (\bibinfo{publisher}{Oxford University Press},
  \bibinfo{year}{2010}).

\bibitem[{\citenamefont{Misawa et~al.}(2006)\citenamefont{Misawa, Yamaji, and
  Imada}}]{misawatcp}
\bibinfo{author}{\bibfnamefont{T.}~\bibnamefont{Misawa}},
  \bibinfo{author}{\bibfnamefont{Y.}~\bibnamefont{Yamaji}}, \bibnamefont{and}
  \bibinfo{author}{\bibfnamefont{M.}~\bibnamefont{Imada}}, \bibinfo{journal}{J.
  Phys. Soc. Jpn.} \textbf{\bibinfo{volume}{75}}, \bibinfo{pages}{064705}
  (\bibinfo{year}{2006}).

\bibitem[{\citenamefont{Schmalian and Turlakov}(2004)}]{schmalian}
\bibinfo{author}{\bibfnamefont{J.}~\bibnamefont{Schmalian}} \bibnamefont{and}
  \bibinfo{author}{\bibfnamefont{M.}~\bibnamefont{Turlakov}},
  \bibinfo{journal}{Phys. Rev. Lett.} \textbf{\bibinfo{volume}{93}},
  \bibinfo{pages}{036405} (\bibinfo{year}{2004}).

\bibitem[{\citenamefont{Green et~al.}(2005{\natexlab{b}})\citenamefont{Green,
  Grigera, Borzi, Mackenzie, Perry, and Simons}}]{Green}
\bibinfo{author}{\bibfnamefont{A.~G.} \bibnamefont{Green}},
  \bibinfo{author}{\bibfnamefont{S.~A.} \bibnamefont{Grigera}},
  \bibinfo{author}{\bibfnamefont{R.~A.} \bibnamefont{Borzi}},
  \bibinfo{author}{\bibfnamefont{A.~P.} \bibnamefont{Mackenzie}},
  \bibinfo{author}{\bibfnamefont{R.~S.} \bibnamefont{Perry}}, \bibnamefont{and}
  \bibinfo{author}{\bibfnamefont{B.~D.} \bibnamefont{Simons}},
  \bibinfo{journal}{Phys. Rev. Lett.} \textbf{\bibinfo{volume}{95}},
  \bibinfo{pages}{086402} (\bibinfo{year}{2005}{\natexlab{b}}).

\bibitem[{\citenamefont{Jakubczyk et~al.}(2010)\citenamefont{Jakubczyk, Bauer,
  and Metzner}}]{Jakub}
\bibinfo{author}{\bibfnamefont{P.}~\bibnamefont{Jakubczyk}},
  \bibinfo{author}{\bibfnamefont{J.}~\bibnamefont{Bauer}}, \bibnamefont{and}
  \bibinfo{author}{\bibfnamefont{W.}~\bibnamefont{Metzner}},
  \bibinfo{journal}{Phys. Rev. B} \textbf{\bibinfo{volume}{82}},
  \bibinfo{pages}{045103} (\bibinfo{year}{2010}).

\bibitem[{\citenamefont{Lukierska-Walasek}(1994)}]{lukierska1994}
\bibinfo{author}{\bibfnamefont{K.}~\bibnamefont{Lukierska-Walasek}},
  \bibinfo{journal}{Acta. Phys. Pol. A} \textbf{\bibinfo{volume}{85}},
  \bibinfo{pages}{381} (\bibinfo{year}{1994}).

\bibitem[{\citenamefont{Carvalho and Plascak}(2015)}]{BEGQTCP}
\bibinfo{author}{\bibfnamefont{D.}~\bibnamefont{Carvalho}} \bibnamefont{and}
  \bibinfo{author}{\bibfnamefont{J.}~\bibnamefont{Plascak}},
  \bibinfo{journal}{Physica A} \textbf{\bibinfo{volume}{432}},
  \bibinfo{pages}{240} (\bibinfo{year}{2015}).

\bibitem[{\citenamefont{Mercaldo et~al.}(2011)\citenamefont{Mercaldo, Rabuffo,
  Naddeo, Caramico~DAuria, and De~Cesare}}]{RGQTCP}
\bibinfo{author}{\bibfnamefont{M.}~\bibnamefont{Mercaldo}},
  \bibinfo{author}{\bibfnamefont{I.}~\bibnamefont{Rabuffo}},
  \bibinfo{author}{\bibfnamefont{A.}~\bibnamefont{Naddeo}},
  \bibinfo{author}{\bibfnamefont{A.}~\bibnamefont{Caramico~DAuria}},
  \bibnamefont{and}
  \bibinfo{author}{\bibfnamefont{L.}~\bibnamefont{De~Cesare}},
  \bibinfo{journal}{EPJ B} \textbf{\bibinfo{volume}{84}}, \bibinfo{pages}{371}
  (\bibinfo{year}{2011}).

\bibitem[{\citenamefont{Kato et~al.}(2014)\citenamefont{Kato, Yamamoto, and
  Danshita}}]{katoQTCP}
\bibinfo{author}{\bibfnamefont{Y.}~\bibnamefont{Kato}},
  \bibinfo{author}{\bibfnamefont{D.}~\bibnamefont{Yamamoto}}, \bibnamefont{and}
  \bibinfo{author}{\bibfnamefont{I.}~\bibnamefont{Danshita}},
  \bibinfo{journal}{Phys. Rev. Lett.} \textbf{\bibinfo{volume}{112}},
  \bibinfo{pages}{055301} (\bibinfo{year}{2014}).

\bibitem[{\citenamefont{Kincaid and Cohen}(1975)}]{kincaid1975phase}
\bibinfo{author}{\bibfnamefont{J.~M.} \bibnamefont{Kincaid}} \bibnamefont{and}
  \bibinfo{author}{\bibfnamefont{E.~G.~D.} \bibnamefont{Cohen}},
  \bibinfo{journal}{Physics Reports} \textbf{\bibinfo{volume}{22}},
  \bibinfo{pages}{57} (\bibinfo{year}{1975}).

\bibitem[{\citenamefont{Kawashima and Harada}(2004)}]{kawashima2004}
\bibinfo{author}{\bibfnamefont{N.}~\bibnamefont{Kawashima}} \bibnamefont{and}
  \bibinfo{author}{\bibfnamefont{K.}~\bibnamefont{Harada}},
  \bibinfo{journal}{J. Phys. Soc. Jpn.} \textbf{\bibinfo{volume}{73}},
  \bibinfo{pages}{1379} (\bibinfo{year}{2004}).

\bibitem[{\citenamefont{Evertz et~al.}(1993)\citenamefont{Evertz, Lana, and
  Marcu}}]{evertz1993}
\bibinfo{author}{\bibfnamefont{H.~G.} \bibnamefont{Evertz}},
  \bibinfo{author}{\bibfnamefont{G.}~\bibnamefont{Lana}}, \bibnamefont{and}
  \bibinfo{author}{\bibfnamefont{M.}~\bibnamefont{Marcu}},
  \bibinfo{journal}{Phys. Rev. Lett.} \textbf{\bibinfo{volume}{70}},
  \bibinfo{pages}{875} (\bibinfo{year}{1993}).

\bibitem[{\citenamefont{Kato and Kawashima}(2010)}]{kato2010}
\bibinfo{author}{\bibfnamefont{Y.}~\bibnamefont{Kato}} \bibnamefont{and}
  \bibinfo{author}{\bibfnamefont{N.}~\bibnamefont{Kawashima}},
  \bibinfo{journal}{Phys. Rev. E} \textbf{\bibinfo{volume}{81}},
  \bibinfo{pages}{011123} (\bibinfo{year}{2010}).

\bibitem[{\citenamefont{Z\"ulicke and Millis}(1995)}]{zulicke}
\bibinfo{author}{\bibfnamefont{U.}~\bibnamefont{Z\"ulicke}} \bibnamefont{and}
  \bibinfo{author}{\bibfnamefont{A.~J.} \bibnamefont{Millis}},
  \bibinfo{journal}{Phys. Rev. B} \textbf{\bibinfo{volume}{51}},
  \bibinfo{pages}{8996} (\bibinfo{year}{1995}).

\bibitem[{\citenamefont{Harada}(2011)}]{harada2011}
\bibinfo{author}{\bibfnamefont{K.}~\bibnamefont{Harada}},
  \bibinfo{journal}{Phys. Rev. E} \textbf{\bibinfo{volume}{84}},
  \bibinfo{pages}{056704} (\bibinfo{year}{2011}).

\bibitem[{\citenamefont{Pelissetto and Vicari}(2002)}]{pelissetto2002}
\bibinfo{author}{\bibfnamefont{A.}~\bibnamefont{Pelissetto}} \bibnamefont{and}
  \bibinfo{author}{\bibfnamefont{E.}~\bibnamefont{Vicari}},
  \bibinfo{journal}{Phys. Rep.} \textbf{\bibinfo{volume}{368}},
  \bibinfo{pages}{549} (\bibinfo{year}{2002}).

\bibitem[{\citenamefont{Emery}(1975)}]{Emery1975}
\bibinfo{author}{\bibfnamefont{V.~J.} \bibnamefont{Emery}},
  \bibinfo{journal}{Phys. Rev. B} \textbf{\bibinfo{volume}{11}},
  \bibinfo{pages}{3397} (\bibinfo{year}{1975}).

\bibitem[{\citenamefont{Furman and Blume}(1974)}]{Blume1974}
\bibinfo{author}{\bibfnamefont{D.}~\bibnamefont{Furman}} \bibnamefont{and}
  \bibinfo{author}{\bibfnamefont{M.}~\bibnamefont{Blume}},
  \bibinfo{journal}{Phys. Rev. B} \textbf{\bibinfo{volume}{10}},
  \bibinfo{pages}{2068} (\bibinfo{year}{1974}).

\bibitem[{\citenamefont{de~Azevedo et~al.}(1995)\citenamefont{de~Azevedo,
  Binek, Kushauer, Kleemann, and Bertrand}}]{de1995}
\bibinfo{author}{\bibfnamefont{M.}~\bibnamefont{de~Azevedo}},
  \bibinfo{author}{\bibfnamefont{C.}~\bibnamefont{Binek}},
  \bibinfo{author}{\bibfnamefont{J.}~\bibnamefont{Kushauer}},
  \bibinfo{author}{\bibfnamefont{W.}~\bibnamefont{Kleemann}}, \bibnamefont{and}
  \bibinfo{author}{\bibfnamefont{D.}~\bibnamefont{Bertrand}},
  \bibinfo{journal}{J. Magn. Magn. Mater.} \textbf{\bibinfo{volume}{140}},
  \bibinfo{pages}{1557} (\bibinfo{year}{1995}).

\bibitem[{\citenamefont{Nakatsuji et~al.}(2008)\citenamefont{Nakatsuji, Kuga,
  Machida, Tayama, Sakakibara, Karaki, Ishimoto, Yonezawa, Maeno, Pearson
  et~al.}}]{YbAlB42008}
\bibinfo{author}{\bibfnamefont{S.}~\bibnamefont{Nakatsuji}},
  \bibinfo{author}{\bibfnamefont{K.}~\bibnamefont{Kuga}},
  \bibinfo{author}{\bibfnamefont{Y.}~\bibnamefont{Machida}},
  \bibinfo{author}{\bibfnamefont{T.}~\bibnamefont{Tayama}},
  \bibinfo{author}{\bibfnamefont{T.}~\bibnamefont{Sakakibara}},
  \bibinfo{author}{\bibfnamefont{Y.}~\bibnamefont{Karaki}},
  \bibinfo{author}{\bibfnamefont{H.}~\bibnamefont{Ishimoto}},
  \bibinfo{author}{\bibfnamefont{S.}~\bibnamefont{Yonezawa}},
  \bibinfo{author}{\bibfnamefont{Y.}~\bibnamefont{Maeno}},
  \bibinfo{author}{\bibfnamefont{E.}~\bibnamefont{Pearson}},
  \bibnamefont{et~al.}, \bibinfo{journal}{Nat.~Phys.}
  \textbf{\bibinfo{volume}{4}}, \bibinfo{pages}{603} (\bibinfo{year}{2008}).

\bibitem[{\citenamefont{Matsumoto et~al.}(2011)\citenamefont{Matsumoto,
  Nakatsuji, Kuga, Karaki, Horie, Shimura, Sakakibara, Nevidomskyy, and
  Coleman}}]{YbAlB42011}
\bibinfo{author}{\bibfnamefont{Y.}~\bibnamefont{Matsumoto}},
  \bibinfo{author}{\bibfnamefont{S.}~\bibnamefont{Nakatsuji}},
  \bibinfo{author}{\bibfnamefont{K.}~\bibnamefont{Kuga}},
  \bibinfo{author}{\bibfnamefont{Y.}~\bibnamefont{Karaki}},
  \bibinfo{author}{\bibfnamefont{N.}~\bibnamefont{Horie}},
  \bibinfo{author}{\bibfnamefont{Y.}~\bibnamefont{Shimura}},
  \bibinfo{author}{\bibfnamefont{T.}~\bibnamefont{Sakakibara}},
  \bibinfo{author}{\bibfnamefont{A.~H.} \bibnamefont{Nevidomskyy}},
  \bibnamefont{and} \bibinfo{author}{\bibfnamefont{P.}~\bibnamefont{Coleman}},
  \bibinfo{journal}{Science} \textbf{\bibinfo{volume}{331}},
  \bibinfo{pages}{316} (\bibinfo{year}{2011}).

\bibitem[{\citenamefont{Deguchi et~al.}(2012)\citenamefont{Deguchi, Matsukawa,
  Sato, Hattori, Ishida, Takakura, and Ishimasa}}]{deguchi}
\bibinfo{author}{\bibfnamefont{K.}~\bibnamefont{Deguchi}},
  \bibinfo{author}{\bibfnamefont{S.}~\bibnamefont{Matsukawa}},
  \bibinfo{author}{\bibfnamefont{N.~K.} \bibnamefont{Sato}},
  \bibinfo{author}{\bibfnamefont{T.}~\bibnamefont{Hattori}},
  \bibinfo{author}{\bibfnamefont{K.}~\bibnamefont{Ishida}},
  \bibinfo{author}{\bibfnamefont{H.}~\bibnamefont{Takakura}}, \bibnamefont{and}
  \bibinfo{author}{\bibfnamefont{T.}~\bibnamefont{Ishimasa}},
  \bibinfo{journal}{Nat. Mater.} \textbf{\bibinfo{volume}{11}},
  \bibinfo{pages}{1013} (\bibinfo{year}{2012}).

\bibitem[{\citenamefont{Watanabe and Miyake}(2010)}]{valence}
\bibinfo{author}{\bibfnamefont{S.}~\bibnamefont{Watanabe}} \bibnamefont{and}
  \bibinfo{author}{\bibfnamefont{K.}~\bibnamefont{Miyake}},
  \bibinfo{journal}{Phys. Rev. Lett.} \textbf{\bibinfo{volume}{105}},
  \bibinfo{pages}{186403} (\bibinfo{year}{2010}).

\bibitem[{\citenamefont{Ramires et~al.}(2012)\citenamefont{Ramires, Coleman,
  Nevidomskyy, and Tsvelik}}]{nodal}
\bibinfo{author}{\bibfnamefont{A.}~\bibnamefont{Ramires}},
  \bibinfo{author}{\bibfnamefont{P.}~\bibnamefont{Coleman}},
  \bibinfo{author}{\bibfnamefont{A.~H.} \bibnamefont{Nevidomskyy}},
  \bibnamefont{and} \bibinfo{author}{\bibfnamefont{A.~M.}
  \bibnamefont{Tsvelik}}, \bibinfo{journal}{Phys. Rev. Lett.}
  \textbf{\bibinfo{volume}{109}}, \bibinfo{pages}{176404}
  (\bibinfo{year}{2012}).

\end{thebibliography}

\end{document}